\documentclass[11pt,showpacs,preprintnumbers,floatfix]{article}
\pdfoutput=1
\usepackage{graphicx} 
\usepackage[colorlinks=true,linkcolor=blue,citecolor=blue,urlcolor=blue,linktocpage=true]{hyperref}
\usepackage{dcolumn} 
\usepackage{bm} 
\usepackage{amsfonts,amsthm,amsmath,amssymb}
\usepackage{epsfig}
\usepackage{color}
\usepackage{textcomp}
\usepackage{hyperref}
\usepackage{titlesec}
\usepackage{slashed}
\usepackage{caption}
\usepackage{subcaption}
\usepackage{siunitx}
\usepackage{booktabs}
\usepackage{multirow}
\usepackage{cite}
\usepackage{placeins}
\usepackage{comment}
\usepackage{braket}
\usepackage{latexsym}

\def\be{\begin{equation}}
\def\ee{\end{equation}}
\def\bea{\begin{eqnarray}}
\def\eea{\end{eqnarray}}

\def\bdm{\begin{displaymath}}
\def\edm{\end{displaymath}}

\def\bq{\begin{quote}}
\def\eq{\end{quote}}

 at 10truept

\newcommand{\bi}{\begin{itemize}}
\newcommand{\ei}{\end{itemize}}

\newcommand{\beq}{\begin{equation}}
\newcommand{\eeq}{\end{equation}}
\newcommand{\p}{\partial}
\newcommand{\ep}{\epsilon}
\newcommand{\lle}{\left \langle}
\newcommand{\rgr}{\right>}
\newcommand{\lb}{\left|}
\newcommand{\rb}{\right|}
\newcommand{\Torder}{\mathrm{T}}
\newcommand{\vect}[1]{\bm{\mathrm{{#1}}}}

\newcommand{\pa}{\partial}

\begin{document}

\vspace*{1cm}

\begin{center}
{\LARGE \bf On the primordial correlation of gravitons with}\\
\vspace{.2cm}
{\LARGE \bf gauge fields}

\vspace*{1.0cm} 
{\large Rajeev Kumar Jain$^{a}$\footnote{\tt rkjain@iisc.ac.in}},
{\large P. Jishnu Sai$^{a}$\footnote{\tt jishnup@iisc.ac.in}}, 
{\large Martin S. Sloth$^{b}$\footnote{\tt sloth@cp3.sdu.dk}}\\
\vspace{.5cm} {\em $^a$Department of Physics, Indian Institute of Science,\\
Bangalore 560012, India}\\
\vspace{.5cm} {\em $^b$CP$^3$-Origins, Centre for Cosmology and Particle Physics Phenomenology, \\ University of Southern Denmark, Campusvej 55, 5230 Odense M, Denmark}\\
\end{center}
\vspace{10pt} 
\begin{abstract}
\noindent
We calculate the primordial correlation of gravitons with an abelian gauge field non-minimally coupled through a dynamical dilaton field or a volume moduli during inflation in the early universe. In particular, we compute the cross-correlation of a tensor mode with two gauge field modes and the corresponding correlation functions for the associated magnetic and electric fields using the in-in formalism. Moreover, using semi-classical methods, we show that the three-point cross-correlation functions satisfy new consistency relations (soft theorems) in the squeezed limit. Our findings exhibit a complete agreement of the full in-in results with the new consistency relations. 
An interesting consequence of our scenario is the possibility of a novel correlation of the primordial tensor mode with the primordial curvature perturbation induced by higher order quantum gravity corrections. The anisotropic background created by long wavelength gauge field modes makes this correlation function non-vanishing. Finally, we discuss how these three-point correlation functions are imprinted on cosmological observables today and the applications to scenarios of inflationary magnetogenesis.
\end{abstract}

\newpage
\tableofcontents

\section{Introduction}
In effective field theories with extra dimensions, as derived from string theory, it is natural to expect that gauge fields are non-minimally coupled in the early universe through a dynamical dilaton field or the moduli of the internal dimensions. This breaks the conformal invariance of the gauge field action and enables the amplification of the gauge fields during inflation. If gauge fields are enhanced in the early universe, during the inflationary epoch, it opens the possibility that such a primordial gauge field could have left a detectable imprint on the observable universe today. It is well known that observable primordial gravitational waves may also be created during inflation by the same causal mechanism, and we therefore expect a non-vanishing non-Gaussian cross-correlation between primordial gravitational waves and primordial gauge fields, which we will calculate here.  Focussing for simplicity on abelian gauge fields, we will consider a model where conformal invariance is broken by a non-minimal coupling between an evolving scalar field and the kinetic term of the gauge field \cite{Turner:1987bw, Ratra:1991bn}, which takes the form
\beq\label{Action1}
\mathcal{L}_{\textrm{gauge}} = -\frac{1}{4}\lambda(\phi) F_{\mu\nu}F^{\mu\nu}~.
\eeq
Such models are often referred to as the kinetic coupling models. There could also be a non-minimal coupling with the parity violating term of the gauge field. However, we shall not consider those models in this paper. In all these scenarios of broken conformal symmetry, one also assumes that the conformal invariance is restored by the end of inflation (or by the end of the reheating epoch). Moreover, such a model only breaks the conformal invariance, and not the gauge invariance. 

 A further motivation for studying this type of models stems from the fact that astrophysical observations indicate the presence of coherent magnetic fields of femto-Gauss strength on large cosmological scales (Mpc scales or larger) exceeding the scales of galaxies and galaxy clusters \cite{Tavecchio:2010mk,Neronov:1900zz,Chen:2014rsa}. It is unclear if any astrophysical process can produce magnetic fields with coherence lengths of cosmological sizes, and it has therefore been speculated that such fields must have an inflationary origin. After all, inflation is known to be the perfect mechanism for producing the very large, even super-horizon,  correlations of primordial fluctuations, needed to explain the observed large scale correlations in the cosmic microwave background (CMB) anisotropies. 

For inflationary magnetogenesis to work, it requires a breaking of the conformal invariance of electromagnetism during inflation, and therefore imply a departure from standard electromagnetism as in eq.~(\ref{Action1})\cite{Ratra:1991bn,Turner:1987bw,Widrow:2002ud}. But even then, back-reaction arising due to the generated electromagnetic fields and issues with entering regimes of strong coupling severely restricts the possibility for significant magnetic fields to be generated during inflation \cite{Demozzi:2009fu}. Only in some very special cases, it is possible for the observed magnetic fields to have been generated during inflation \cite{Durrer:2010mq, Byrnes:2011aa, Jain:2012jy, Ferreira:2013sqa,Ferreira:2014hma,Kobayashi:2019uqs}. It would therefore be very important to understand if magnetic fields observed on cosmological scales really have an inflationary origin or not.

One possibility is to look at the correlation of the cosmological magnetic fields with the inflaton perturbation, or equivalently the gauge-invariant curvature perturbation. Such correlations have earlier been discussed in \cite{Caldwell:2011ra, Motta:2012rn, Jain:2012ga,Jain:2012vm,Biagetti:2013qqa}, and also in \cite{Chowdhury:2018blx,Chowdhury:2018mhj} in a different set-up. The idea being that magnetic fields, if they are indeed generated during inflation, will be correlated with the primordial curvature perturbation. On the other hand, if they are generated after inflation, no such correlation will exist. Assuming an exact model for generating the magnetic fields during inflation, one can compute the expected correlation. Since one of the few models, which can indeed generate significant primordial magnetic fields during inflation, takes the general form in eq.\,(\ref{Action1}), much attention was focussed on this model in the previous work, and in \cite{Jain:2012ga,Jain:2012vm} the non-Gaussian correlation function of the curvature perturbation with the magnetic field
\beq
\left< \zeta(\bf{k_1}) B(\bf{k_2})\cdot  B(\bf{k_3})\right>~,
\eeq
was calculated. It was shown that such a non-Gaussian correlation function satisfies a new consistency relation in the squeezed limit and its magnitude becomes quite large in the flattened configuration in the Fourier space. Later, some consistency relations for the soft limit of the higher order correlators involving magnetic fields and matter over-densities were proposed and it was pointed out that any violation of such consistency relations would point towards an inflationary origin of cosmological magnetic fields \cite{Berger:2014wta}.

Here we will consider the natural extension of that work, and look at the non-trivial cross correlations of primordial gauge fields and primordial gravitational waves.  Our main objective is therefore to compute the primordial non-Gaussian correlations functions
\beq \label{ngcorr}
\left< \gamma(\bf{k_1}) A(\bf{k_2})\cdot  A(\bf{k_3})\right>\, ,\qquad  \left< \gamma(\bf{k_1}) B(\bf{k_2})\cdot  B(\bf{k_3})\right>\, ,\qquad   \left< \gamma(\bf{k_1}) E(\bf{k_2})\cdot  E(\bf{k_3})\right>~,
\eeq     
where $\gamma$ is the tensor mode, ${\bf A}$ is our abelian gauge boson (vector potential) and ${\bf B}$,  ${\bf E}$ are the associated magnetic and electric fields, respectively. Since these are equal time correlation functions, we will calculate them using the full in-in formalism which is widely used to calculate such higher order correlators during inflation \cite{Giddings:2010ui}. 

As a way of checking our final results, we propose new semi-classical consistency relations (soft theorems\footnote{Since the semi-classical relations relate a higher order correlation function involving an infrared (soft mode) to a lower order correlation function, the semi-classical relations are also sometimes called the soft theorems with a reference to the Weinberg's soft photon and graviton theorems \cite{Weinberg:1965nx}, which can be related to the asymptotic symmetries and the gravitational memory of flat space \cite{Strominger:2014pwa,Strominger:2017zoo}. For a discussion of the relation between semi-classical relations/soft theorems, asymptotic symmetries and gravitational memory in inflation, see \cite{Ferreira:2016hee,Ferreira:2017ogo,Ferreira:2017erz} and also \cite{Hui:2018cag,Hui:2018cag,Kehagias:2017rpe,Hinterbichler:2016pzn,Hinterbichler:2012nm,Hinterbichler:2013dpa}.}) in the limit wherein the wavelength of the graviton goes to infinity (the squeezed limit). The semi-classical consistency relation for correlation functions only involving the curvature perturbation, $\zeta$,  or the tensor mode, $\gamma$, was first proposed by Maldacena in \cite{Maldacena:2002vr, Maldacena:2011nz} (and also subsequently by others \cite{Creminelli:2004yq, Cheung:2007sv}), hence they are also sometimes called the Maldacena consistency relations. A related consistency relation for gauge fields was first discussed in  \cite{Jain:2012ga}, wherein a {\it{magnetic}} consistency relation for the correlation function $\left< \zeta\, {\bf B}\cdot {\bf B}\right>$ was proposed. Here we will consider the natural extension to the correlation functions above in eq.~(\ref{ngcorr}), and show that the full in-in correlation function in fact satisfies the semi-classical consistency relation in the appropriate limits\footnote{The semi-classical relations serve an important role as a cross check of the correctness of the full in-in correlation functions (not only Maldacena's three-point function). As was first shown in \cite{Seery:2008ax}, it can also be used to calculate the contribution to the four-point function in the limit where the momentum of the exchanged graviton goes to zero, the so called counter collinear limit, by simply taking the square of the three-point function in the squeezed limit. In \cite{Giddings:2010nc}, it was further realized that one can also extract the infrared divergent part of higher order loop diagrams using such semi-classical relations.}. 

One of the problems with generating large magnetic fields during inflation, is that the universe becomes an almost perfect conductor after reheating, which very rapidly erases any magnetic field which would have been generated during inflation. Models of successful magnetogenesis therefore typically require a very low scale of inflation  \cite{Ferreira:2013sqa,Ferreira:2014hma}, or at least a very low reheating temperature \cite{Kobayashi:2019uqs}.
This problem is absent if the magnetic field belongs to a dark gauge group. Therefore we can more easily imagine that large dark magnetic fields could have been generated during inflation. While such dark magnetic fields are not directly observable in the same way as the standard magnetic fields of electromagnetism, they do leave indirect imprints in the curvature perturbation of the universe. The total curvature perturbations is then the sum of the "usual" contribution, $\zeta_0$ generated independently of the magnetic field, and the contribution, $\zeta_B$, imprinted by magnetic fields
\beq
\zeta(\tau) \simeq \zeta_0(\tau) + \zeta_B(\tau)~,
\eeq
where the curvature perturbation induced by any magnetic field is \cite{Nurmi:2013gpa}
\beq
\zeta_B(\tau) = \int_{\tau_0}^\tau d\ln \tilde{\tau} \lambda(\tilde{\tau}) \frac{B_i B^i}{3 H^2 \ep}~.
\eeq
This expression is valid for both dark and ordinary magnetic fields. In the remainder of this paper, it can refer to both cases. The magnetic fields generated during inflation follow a Gaussian statistics to the leading order in perturbations which implies that the induced curvature perturbation $\zeta_B$ is generally a non-Gaussian field. Various imprints of this induced component such as non-Gaussianities and anisotropies in the power spectrum and bispectrum have also been greatly studied \cite{Yokoyama:2008xw, Barnaby:2012tk, Soda:2012zm, Bartolo:2012sd, Lyth:2013sha, Abolhasani:2013zya, Shiraishi:2013vja, Fujita:2013qxa}.

One may naively expect that due to the induced curvature perturbations, there might exist a direct non-trivial correlation of the primordial tensor mode with the primordial curvature perturbation of the form
\beq
\left< \gamma \zeta\right> \simeq \left< \gamma \zeta_B\right> \propto \left<\gamma {\bf B\cdot B}\right>,
\eeq
due to the presence of the non-vanishing correlator $\left<\gamma {\bf B\cdot B}\right>$. A similar contribution will arise from the corresponding electric fields as well. As we shall show in section \ref{twocorr}, while $\left< \gamma \zeta_B\right>$ vanishes in the isotropic limit\footnote{Some scaling arguments of such a mixed correlator based on the  special conformal transformations have been discussed in \cite{BeltranAlmeida:2019gku} wherein there is an explicit parity violating term in the Lagrangian.}, such a two point correlation function can actually be non-vanishing in the anisotropic background created by the long wavelength gauge fields, and may even receive contributions of the form $\left<\gamma {\bf B\cdot B}\right>$ from quantum gravity. 

This paper is organized as follows. In the next section we will quickly review the dynamics of tensor perturbations during inflation and the kinetic coupling model as a mechanism for the production of large scale primordial gauge fields during inflation. In section \ref{bispectrum}, We  calculate the full bispectrum associated with the cross correlations of the inflationary tensor perturbation with gauge fields using the very general in-in formalism and discuss the extent of non-linearities using the Fourier space shape functions.  In section \ref{semi-classical}, we derive the consistency relations for these non-Gaussian cross correlations using a semi-classical approach.  In section \ref{twocorr}, we calculate a direct correlation of the primordial tensor mode and the primordial curvature perturbation induced by the gauge fields in the anisotropic background of long wavelength modes. Finally, in section \ref{conclusions}, we summarise our results and conclude with a discussion. In appendix \ref{integrals}, we have listed some useful integrals appearing in the in-in results of various cross correlations. In appendix \ref{validity-sca}, we discuss the validity of the semi-classical approach used to derive our consistency relations and in appendix \ref{dynamical}, we discuss the presence of the dynamical correction terms arising in the cross correlation involving electric fields in the in-in formalism.

Throughout this paper, we work in natural units with $\hbar = c =1$, and the Planck mass $M_{\rm Pl}^2 =1/8\pi G$ is set to unity. Our metric convention is $(-,+,+,+)$.

\section{Dynamics of primordial tensors and gauge fields during inflation}
\label{dynamics-tensor-gauge}

In this section, we shall briefly discuss the dynamical evolution of primordial tensor modes and abelian gauge fields during inflation.  The homogeneous and isotropic background during the inflationary expansion is described by the spatially flat FLRW metric
\bea
ds^2 = -dt^2 + a^2(t)\, d{\bf x}^2 = a^2(\tau)\left(-d\tau^2+d{\bf x}^2\right),
\eea
where $\tau$ is the conformal time, defined by $d\tau = dt/a$ and $a(\tau)$ is the scale factor. 
The perturbed FLRW metric in the presence of tensor modes is given by 
\bea
ds^2 =-dt^2+a^2(t) \bigl[e^{\gamma}\bigr]_{ij}dx^i dx^j \approx -dt^2+a^2(t) \bigl[\delta_{ij} + \gamma_{ij}\bigr]dx^i dx^j~,
\eea
where $\gamma_{ij}$ is the metric tensor perturbation which is transverse and traceless i.e. $\gamma_{{ij},j}=0$ and $\gamma_{ii}=0$. 
These conditions lead to only two independent radiative degrees of freedom of tensor perturbations which correspond to the two polarizations of gravitational waves (GW). 
To the linear order in metric fluctuations, there is typically no active source of GWs during inflation. However, quantum fluctuations of $\gamma_{ij} $ are parametrically amplified by the quasi-exponential expansion of the universe. In order to describe this phenomenon, one needs to quantize the canonical degrees of freedom associated with the tensor perturbations. 
Following the usual quantization formalism, the corresponding mode expansion of tensor perturbation is defined as
\bea
\label{gamma_ij}
\gamma_{ij}({\bf x},\tau) 
= \int\frac{d^{3}{\bf k}}{(2\pi)^{3}}\sum_{s = \pm 2}
\left[\gamma_{k}(\tau) \, e^{i{\bf k}\cdot {\bf x}} \,  \epsilon ^{s}_{ij}(\hat{\bf k}) \, b_{\bf k}^s + h.c. \right],
\eea
where $\epsilon_{ij}^{s}$ is the polarization tensor corresponding to the helicity $s$, with the normalization $\epsilon_{ij}^{s} \epsilon_{ij}^{*s'}=2\delta_{ss'}$ and the creation and annihilation operators do satisfy the condition, $[b_{\bf k}^s,b^{s' \dagger}_{\bf k'}]=(2\pi)^3 \delta^{(3)}({\bf k}-{\bf k'}) \delta_{s s'} $. 
The tensor helicity mode $\gamma^s({\bf k},\tau)$ can be defined through the following equation, 
\bea
\label{g_helicitymode}
\gamma_{ij}({\bf x},\tau) 
= \sum_{s = \pm 2}\int\frac{d^{3}{\bf k}}{(2\pi)^{3}} \left[\epsilon ^{s}_{ij}(\hat{\bf k})
\gamma^s({\bf k},\tau) \, e^{i{\bf k}\cdot {\bf x}}  \right] \,.  
\eea
We can now define the two point function in Fourier space as
\bea
\label{tensor_Pwrspctrm}
\langle\gamma^s({\bf k},\tau)\gamma^{s'}({\bf k'},\tau)\rangle=(2\pi)^3\delta^{(3)} (\vect{k}+\vect{k'})P_{\gamma}(k,\tau)\delta_{ss'}
\eea
where the power spectrum of tensor modes is given by $P_{\gamma}(k,\tau)=|\gamma_k (\tau)|^2$. It is often useful to define a dimensionless tensor power spectrum ${\cal P}_{\gamma}$ as
\beq
\label{ps-dimensionless}
{\cal P}_{\gamma} (k,\tau) = \frac{k^3}{2 \pi^2} P_{\gamma} (k,\tau) = \frac{k^3}{2 \pi^2}|\gamma_k (\tau)|^2,
\eeq
which is usually evaluated in the super-horizon limit $|k\tau| \to 0$.
For a nearly de-Sitter background $\tau \simeq -1/aH$ where $H$ is the Hubble parameter, the Bunch-Davies normalized solution for the mode function $\gamma_{k}(\tau)$ is given by
\beq
\label{modefn_g}
\gamma_k(\tau) = \frac{H}{\sqrt{k^3}}(1+ik\tau)\,e^{-ik\tau} ~,
\eeq
which, in the super-horizon limit, leads to ${\cal P}_{\gamma} \sim (H/2 \pi)^2$. The tensor power spectrum thus measures the energy scale during inflation and is exactly scale invariant in a de-Sitter background (and nearly scale invariant in a slow roll inflation). 

As mentioned already in the introduction, it is well known that gauge fields described by the canonical Maxwell's action are not amplified in the FLRW background due to the conformal symmetry and thus, a necessary condition for amplification of gauge fields during inflation is to break this conformal invariance. One of the simplest possibility which has been explored to a large extent in the literature is the so called kinetic coupling model which breaks the conformal invariance (while preserving gauge invariance) by explicitly coupling gauge fields to the inflaton as \cite{Ratra:1991bn}
\beq\label{A1}
S_{A} = -\frac{1}{4}\int d^4 x \sqrt{-g}\,\lambda(\phi) F_{\mu\nu} F^{\mu\nu} + \int d^4 x \sqrt{-g}\, j^{\mu}A_{\mu},
\eeq
where the gauge field tensor is defined as $F_{\mu\nu} = \pa_\mu A_\nu-\pa_\nu A_\mu$ and the second term represents the interaction
where $j^{\mu}$ is the four current density. For simplicity, we shall neglect the interaction term and assume that there are no free charges. In the Coulomb gauge wherein $A_0=0$ and $\pa_iA^i=0$, the quadratic action for the gauge field $A_i$ becomes
\beq\label{A2}
S_{A} =\frac{1}{2} \int d^3 x\, d\tau \lambda(\phi) \left({A_i'}^2 -\frac{1}{2}(\p_iA_j-\p_jA_i)^2\right).
\eeq
Since we are interested in the dynamical amplification of the quantum gauge field originating from the vacuum, we promote the classical gauge field $A_i$ to an operator and impose the canonical quantization conditions. Following this, we define the usual mode expansion as
\beq
\label{Mode_expnA }
A_{i}({\bf x},\tau) 
= \int \frac{d^{3} {\bf k}}{(2\pi)^{3}} 
\sum_{\lambda = \pm 1} \left[ A_{k}(\tau) \, e^{i{\bf k}\cdot {\bf x}}\, \epsilon ^{(\lambda)}_{i}(\hat{\bf {k}}) \, a_{\bf k}^\lambda + h.c. \right], 
\eeq
and impose the standard commutation relations
\beq
[a_{\bf k}^\lambda,a^{\lambda' \dagger}_{\bf k'}] =(2\pi)^3\delta^{(3)}(\vect{k}-\vect{k}')\delta_{\lambda\lambda'} ~.
\eeq
Here $A_k(\tau)$ is the mode function in Fourier space and the polarization vectors will satisfy the usual relations: $\vect{k}\cdot \ep^{\sigma}(\hat k) =0$, $\ep^{\sigma}(\hat {\bf k})\cdot{\ep^{\sigma'}}^*(\hat {\bf k}) =\delta_{\sigma \sigma'}$, and $\sum_{\sigma=\pm} \ep_i^{\sigma}(\hat {\bf k}){\ep_j^{\sigma}}^*(\hat {\bf k}) =\delta_{ij} - k_i k_j/k^2$. 

One can now define a canonically normalized gauge field $v_i= S(\tau) A_i$ where $S^2(\tau) =\lambda(\phi(\tau))$. In this case, the quadratic action for the gauge field takes the canonical form as
\beq
S_{A} =   \frac{1}{2}\int d^3 x\, d\tau   \left[v_i'^2-(\p_j v_i)^2+\frac{S''}{S}v_i^2\right]~,
\eeq
and the equation of motion for the rescaled mode function, $v_k = S(\tau) A_k$, takes the form of a harmonic oscillator with a time dependent mass term as
\beq
\label{cngf}
v_k'' +\left(k^2-\frac{S''}{S}\right)v_k=0~.
\eeq
In order to compute the two point and higher order correlators, one needs to define the coupling function $\lambda(\phi)$. For instance, the dilatonic coupling corresponds to $\lambda(\phi) \sim {\rm exp}(2\phi)$ \cite{Ratra:1991bn, Martin:2007ue}. Since $\phi$ must be dynamical, it turns out to be more convenient to parametrize the time dependence of the coupling function as $\lambda(\phi(\tau))= \lambda_I(\tau/\tau_I)^{-2n}$ where $\tau_I$ is the conformal time at the end of inflation, and $\lambda_I$ is the coupling evaluated at the end of inflation $\tau = \tau_I$.

For a power law coupling $\lambda(\tau)= \lambda_I(\tau/\tau_I)^{-2n}$, one finds that $S''/S=(\nu^2-1/4)/\tau^2$ with $\nu=(n+1/2)$ and the mode solution can be written in terms of the Hankel functions.
In the sub-horizon limit, the mode solution must be normalized to the Bunch-Davies vacuum
\beq
v_k(\tau) = \frac{1}{\sqrt{2k}}e^{-i k \tau}~,
\eeq
which leads to the full solution for $v_k$ as
\bea\label{vk}
v_k(\tau) = \frac{\sqrt{\pi}}{2}e^{i\pi(n+1)/2}\sqrt{-\tau}H^{(1)}_{n+\frac{1}{2}}(-k\tau)~.
\eea
With this solution, the mode function $A_k(\tau)$ is expressed as
\bea
A_k(\tau) = \frac{1}{\sqrt{\lambda_I}}\frac{\sqrt{\pi}}{2}e^{i\pi(n+1)/2}\sqrt{-\tau}\left(\frac{\tau}{\tau_I}\right)^nH^{(1)}_{n+\frac{1}{2}}(-k\tau)~, \label{modfn_A}
\eea
where $H_{n}^{(1)}(x)$ is a Hankel function of the first kind. The two point correlation function of gauge fields $A_i$ can be obtained as follows, 
\beq
\left< A_i(\tau,\vect{k}) A_j(\tau,\vect{k}') \right>= (2\pi)^3 \delta^{(3)}(\vect{k}+\vect{k}')\left(\delta_{ij}-\frac{ k_i  k_j}{k^2}\right) \left| A_k(\tau)\right|^2.
\eeq
With respect to an observer characterised by the 4-velocity vector $u^{\mu}$, one can covariantly define the electric field $E_\mu $  and magnetic field $B_\mu$ as  \cite{Subramanian:2009fu}.
\begin{equation}
E_\mu =F_{\mu \nu }u^{\nu} , \qquad
B_\mu = {^*F}_{\mu\nu} u^\nu \,,
\label{ebnudef0}
\end{equation}
where ${^*F}_{\mu\nu}$ is the dual of the electromagnetic field tensor $F_{\mu \nu}$ which is defined by the relation
\begin{equation}
{^*F}_{\mu\nu} =\frac{1}{2} \eta_{\mu\nu\alpha\beta} F^{\alpha\beta},
\end{equation}
with $\eta_{\mu\nu\alpha\beta}$ being the totally antisymmetric permutation tensor of space-time with $\eta_{0123} =\sqrt{-g}$. In this work, the covariant magnetic and electric fields are defined with respect to a comoving observer having 4-velocity $u^{\mu}=(\frac{1}{a},0,0,0)$.
The power spectra of the magnetic field and electric field are defined as
\bea
\label{powerspectra}
\left<B_{\mu}(\tau,\vect{k}) B^{\mu}(\tau,\vect{k}')\right> & = &(2\pi)^3\delta^{(3)} (\vect{k}+\vect{k}') P_B(k, \tau), \\
\left<E_{\mu}(\tau,\vect{k}) E^{\mu}(\tau,\vect{k}')\right> & = &(2\pi)^3\delta^{(3)} (\vect{k}+\vect{k}') P_E(k, \tau).
\eea
Note that, both $P_B$ and $P_E$ are dimension-full. We can now easily express $P_B$ and $P_E$ in terms of the mode function $A_k(\tau)$ as
\bea
\label{P_B}
 P_B(k, \tau)&=&2\frac{k^2}{a^4}|A_k(\tau)|^2 \\
 \label{P_E}
 P_E(k, \tau)&=&\frac{2}{a^4}|A'_k(\tau)|^2
\eea
which can be readily calculated using the obtained mode solution (\ref{modfn_A}) for the case of a power law coupling \cite{Martin:2007ue, Subramanian:2009fu, Caldwell:2011ra, Jain:2012vm}. 
For the cases of $n=2$ and $n=-3$, the magnetic field spectrum turns out to be scale invariant i.e. $P_B \propto k^{-3}$ while the electric field spectrum behave as $P_E \propto k^{-1}$ and $P_E \propto k^{-5}$, respectively. In both these cases, there arise various issues such as the strong coupling and the back reaction problem which have been discussed to a great extent in the literature \cite{Martin:2007ue, Demozzi:2009fu, Subramanian:2009fu, Ferreira:2013sqa, Ferreira:2014hma}.

\section{Cross-correlation of inflationary tensor perturbation with primordial gauge fields}
\label{bispectrum}

In this section, we shall discuss our calculations of the cosmological cross-correlations of primordial tensor perturbation with gauge fields using the in-in formalism while in the next section, we shall discuss the semiclassical approach to arrive at these correlators. As expected, both these approaches lead to the same results for these cross-correlations in the squeezed limit. 


\subsection{A complete calculation using in-in formalism}
In order to compute the correlation function during inflation, we adopt a very powerful tool of the in-in formalism which can be used to calculate various correlation functions at equal times. A detailed discusses of the in-in formalism can be found in appendix \ref{dynamical}. In this formalism, the expectation value of an operator $\mathcal{O}$ at time $\tau_I$ is given by
\begin{equation}
    \label{exp1_in-in}
    \lle \mathcal{O}(\tau_I)  \rgr = \lle
    0\rb
    {\bar \Torder}\left(e^{i\int_{-\infty}^{\tau_I}d\tau H_{\rm int}}\right) \mathcal{O}(\tau_I) \Torder \left(e^{-i\int_{-\infty}^{\tau_I} d\tau H_{\rm int}}\right)\lb
    0 \rgr
\end{equation}
where $H_{\rm int}$ is the interaction Hamiltonian and $\Torder$, $\bar \Torder$ are the time ordering and anti-time ordering operators, respectively. The quantum state $\lb 0 \rgr$ refers to the physical vacuum state of the theory which we take to be the standard Bunch-Davies vacuum state. We shall first apply the in-in formalism to calculate the cross-correlations of the tensor mode with gauge fields.
\subsubsection{$\langle \gamma A_{\mu}A^{\mu}\rangle$} \label{Sec_gAA}

Let us first compute the cross-correlation of a tensor mode with two gauge field modes, i.e., a correlator of the form $\langle \gamma A_{\mu}A^{\mu}\rangle$. We first find that
\bea
\label{gAA_expnsn}
 \gamma A_{\mu}A^{\mu}=\frac{1}{a^2}\gamma A_iA_i -\frac{1}{a^2}  \gamma \gamma_{ij}A_iA_j + {\cal O}(\gamma^3)
\eea
where all the repeated indices in the above and following expressions are summed over (irrespective of up or down). By using  the master formula of the in-in formalism (\ref{exp1_in-in}), we can calculate $\langle \gamma A_{\mu}A^{\mu}\rangle$ perturbatively which, to the leading order, is given by
\bea
\label{pertb_gAA}
\langle\gamma({\bf k}_1,\tau_I) A_{\mu}({\bf k}_2,\tau_I)A^{\mu}({\bf k}_3,\tau_I)\rangle &=& \lle  0\rb \gamma({\bf k}_1,\tau_I) A_{\mu}({\bf k}_2,\tau_I)A^{\mu}({\bf k}_3,\tau_I) \lb  0 \rgr \nonumber \\ &+& i\int_{-\infty}^{\tau_I}d\tau \lle  0\rb[H_{\rm int}(\tau),\gamma({\bf k}_1,\tau_I) A_{\mu}({\bf k}_2,\tau_I)A^{\mu}({\bf k}_3,\tau_I)]\lb  0 \rgr,
\eea
where $H_{\rm int}$ is the cubic order interaction Hamiltonian which can be calculated from equation (\ref{A1}),
\begin{equation}
\label{H_gAA}
             H_{\rm int}(\tau)=\frac{1}{2} \int d^3x \, \lambda(\tau) \left(\gamma^{ij}A'_iA'_j-\gamma^{ij}\delta^{kl}(\partial_iA_k\partial_jA_l+\partial_kA_i\partial_lA_j)+2\gamma^{ij}\delta^{kl}\partial_iA_k\partial_lA_j \right).
         \end{equation}
         \begin{figure}[!t] 
\begin{center}
\includegraphics[width=11cm, height=6cm]{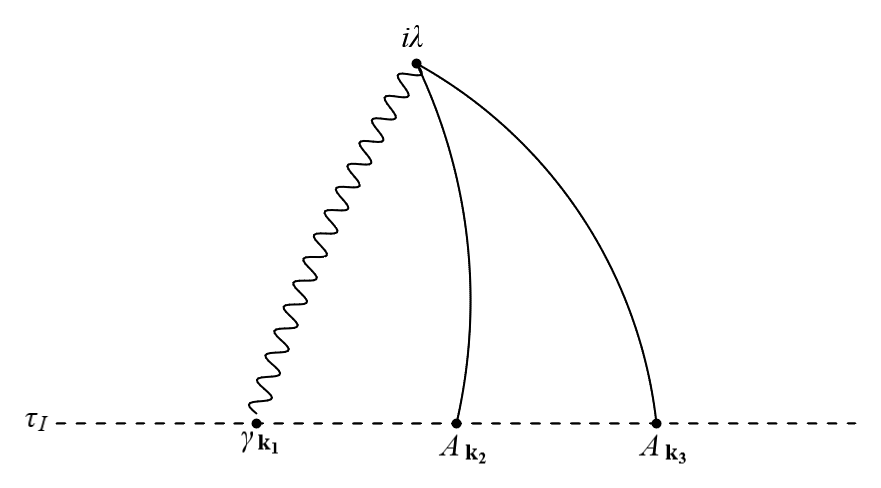}
\end{center}
\caption{The leading order cosmological diagram contributing to all equal time $\tau_I$ cross-correlations of our interest. In addition to the contribution from this diagram, there will also be additional corrections for each correlator, as discussed in the text.  For  $\lle \gamma A_{\mu}A^{\mu} \rgr$ and $\lle \gamma B_{\mu}B^{\mu} \rgr$, there is a kinematical correction term. But for the case of $\lle \gamma E_{\mu}E^{\mu} \rgr$, there is not only a kinematical correction term but also a dynamical correction term. }
\label{cosmodiag}
\end{figure} 
By substituting (\ref{gAA_expnsn}) in (\ref{pertb_gAA}), we observe that the first term on the right hand side (RHS) of equation (\ref{pertb_gAA}) gives a non-zero contribution. It can be easily calculated from the Fourier mode functions (\ref{modefn_g}) and (\ref{modfn_A}) as,
\bea
\lle  0\rb \gamma({\bf k}_1,\tau_I) A_{\mu}({\bf k}_2,\tau_I)A^{\mu}({\bf k}_3,\tau_I) \lb  0 \rgr = \frac{1}{2} (2\pi)^3\delta^{(3)} (\vect{k}_1+\vect{k}_2+\vect{k}_3)\epsilon_{ij}\frac{k_{2i}k_{2j}}{\tilde{k}_2^{2}} P_\gamma(k_1,\tau_I) P_A(\tilde{k}_2,\tau_I).
\eea
Here, ${\bf \tilde{k}}_2={\bf k_2}+\frac{1}{2}{\bf k_1}$, $P_\gamma(k,\tau)$ is the power spectrum of the tensor modes, as defined in  (\ref{tensor_Pwrspctrm}) and $P_A(k,\tau)$ is the power spectrum of the gauge fields defined as, 
\bea
\langle A_{\mu}({\bf k},\tau)A^{\mu}({\bf k'},\tau)\rangle=(2\pi)^3\delta^{(3)}(\vect{k}+\vect{k}')P_A(k,\tau).
\eea
Thus, the vacuum expectation value of $\gamma A_{\mu}A^{\mu}$ does contribute to this non-Gaussian correlator at the leading order. Since this contribution is originating purely from the kinematical considerations without using the dynamics of gauge fields, we refer to such contributions as the \textit{kinematical correction terms}. As we shall see later, similar kinematical terms will  also appear for the cases of $\langle \gamma B_{\mu}B^{\mu}\rangle$ and $\langle \gamma E_{\mu}E^{\mu}\rangle$, respectively. 

Now, the second term in the RHS of equation (\ref{pertb_gAA}) is calculated using the cosmological diagrammatic rules \cite{Giddings:2010ui}. The leading order diagram contributing to this correlator is shown in figure \ref{cosmodiag}. Substituting the interaction Hamiltonian $H_{\rm int}$ from equation (\ref{H_gAA}) in equation (\ref{pertb_gAA}) and after some simplifications, we obtain the total contribution as, 
\bea
\label{gammaAA1}
\langle\gamma({\bf k}_1) A_{\mu}({\bf k}_2)A^{\mu}({\bf k}_3)\rangle &=&(2\pi)^3 \delta^{(3)}({\bf k_1}+{\bf k_2}+{\bf k_3})\epsilon_{ij}\Biggl[ \frac{1}{2} \frac{k_{2i}k_{2j}}{\tilde{k}_2^{2}}P_{\gamma}(k_1)P_A(\tilde{k}_2)\nonumber \\ &+&\left(\delta_{mi}-\frac{k_{2m}k_{2i}}{k_2^2}\right)\left(\delta_{mj}-\frac{k_{3m}k_{3j}}{k_3^2}\right)\left(\frac{\mathcal{I}_1}{a^2}+({\bf k_2 \cdot k_3})\frac{\mathcal{I}_2}{a^2}\right) \nonumber \\ &+& k_{2i}k_{3j}\left(1+\frac{({\bf k_2 \cdot k_3})^2}{k_2^2 k_3^2}\right)\frac{\mathcal{I}_2}{a^2} -k_{3i}k_{2n}\left(\delta_{mn}-\frac{k_{3m}k_{3n}}{k_3^2}\right)\left(\delta_{mj}-\frac{k_{2m}k_{2j}}{k_2^2}\right)\frac{\mathcal{I}_2}{a^2} \nonumber \\ &-&k_{2j}k_{3n}\left(\delta_{mn}-\frac{k_{2m}k_{2n}}{k_2^2}\right)\left(\delta_{mi}-\frac{k_{3m}k_{3i}}{k_3^2}\right)\frac{\mathcal{I}_2}{a^2} \Biggr ],
\eea
where the integrals $\mathcal{I}_1$ and $\mathcal{I}_2$ can be expressed in terms of the mode functions of tensor mode and gauge field and their time derivatives as
\bea
\mathcal{I}_1&=& 2\,{\rm Im} \left[\gamma_{k_1}(\tau_I)A_{k_2}(\tau_I)A_{k_3}(\tau_I)\int d\tau  \lambda(\tau) \gamma_{k_1}^*(\tau)A_{k_2}^{'*}(\tau)A_{k_3}^{'*}(\tau)\right]\label{I1}, \\
\mathcal{I}_2 &=& 2\,{\rm Im} \left[\gamma_{k_1}(\tau_I)A_{k_2}(\tau_I)A_{k_3}(\tau_I)\int d\tau  \lambda(\tau) \gamma_{k_1}^*(\tau)A_{k_2}^*(\tau)A_{k_3}^*(\tau)\right]\label{I2},
\eea
and the explicit time dependence, $\tau_I$, of the operators and the power spectra in the equation (\ref{gammaAA1}) is omitted for notational convenience.  
It is useful to rewrite these integrals in (\ref{I1}) and (\ref{I2}) in terms of another set of auxiliary integrals, denoted by $\tilde{\mathcal{I}}_n^{(1)}$ and $\tilde{\mathcal{I}}_n^{(2)}$, as follows 
\bea
\mathcal{I}_1&=&  -|\gamma^{(0)}_{k_1}(\tau_I)|^2 |A^{(0)}_{k_2}(\tau_I)| |A^{(0)}_{k_3}(\tau_I)| ~k_2 k_3~\tilde{\mathcal{I}}_n^{(1)} \label{Int1}, \\
\mathcal{I}_2&=&  -|\gamma^{(0)}_{k_1}(\tau_I)|^2 |A^{(0)}_{k_2}(\tau_I)| |A^{(0)}_{k_3}(\tau_I)| ~ \tilde{\mathcal{I}}_n^{(2)}\label{Int2},
\eea
where $\gamma^{(0)}_{k}(\tau_I)$ and $A^{(0)}_{k}(\tau_I)$ are the asymptotic super-horizon expansion of the mode functions (\ref{modefn_g}) and (\ref{modfn_A}), given by
\bea
|\gamma^{(0)}_{k}(\tau_I)| &=&  \frac{H}{\sqrt{k^3}} \\
|A_k^{(0)}(\tau_I)| &=& \begin{cases}\frac{1}{\sqrt{\lambda_I}}\; \frac{2^n}{\sqrt{2\pi}} \; \frac{\Gamma(n+\frac{1}{2})}{\sqrt{k}}(-k\tau_I)^{-n}, \; \text{for} \; n>-\frac{1}{2}\\  \\
\frac{1}{\sqrt{\lambda_I}}\; \frac{\sqrt{2\pi}}{2^{n+1}\Gamma(n+\frac{3}{2})} \; \frac{1}{\sqrt{k}}\; (-k\tau_I)^{n+1}, \; \text{for}\; n<-\frac{1}{2}
\end{cases}
\eea
and the explicit forms of the integrals $\tilde{\mathcal{I}}_n^{(1)}$ and $\tilde{\mathcal{I}}_n^{(2)}$ corresponding to both $n>-\frac{1}{2}$ and $n<-\frac{1}{2}$ are given in Appendix \ref{integrals}.

One can straightforwardly substitute (\ref{I1}) and (\ref{I2}) into (\ref{gammaAA1}) and after considerable simplification, the final result for the correlator $\langle \gamma A_{\mu}A^{\mu}\rangle$ can be written as
\bea
\label{gAA-final}
\langle\gamma({\bf k}_1) A_{\mu}({\bf k}_2)A^{\mu}({\bf k}_3)\rangle &=& (2\pi)^3 \delta^{(3)}({\bf k_1}+{\bf k_2}+{\bf k_3})\epsilon_{ij}\frac{k_{2i}k_{2j}}{k_2^2}\nonumber \\ &\times &\left[{\cal M}_A(k_1,k_2,k_3,\tau_I){\cal W}_A(k_1,k_2,k_3)+\frac{1}{2} \left(\frac{k_2}{\tilde{k}_2}\right)^2 P_{\gamma}(k_1)P_A(\tilde{k}_2)\right],
\eea
with 
\bea
{\cal M}_A(k_1,k_2,k_3,\tau_I)&=&\frac{|\gamma^{(0)}_{k_1}(\tau_I)|^2 |A^{(0)}_{k_2}(\tau_I)| |A^{(0)}_{k_3}(\tau_I)|}{a^2} \; , \\
{\cal W}_A(k_1,k_2,k_3)&=&\left[({\bf k_2}+{\bf k_3})^2-({\bf k_2} \cdot {\bf k_3})\right]\frac{k_2 \tilde{\mathcal{I}}_n^{(1)}}{k_3} +k_2^2 \tilde{\mathcal{I}}_n^{(2)}.
\eea
Here the subscript `$A$' in ${\cal M}_A$ and $W_A$ refers to the gauge field $A_{\mu}$. Note that, in the final result (\ref{gAA-final}), the first contribution arises from the interaction term corresponding to the leading order cosmological diagram in figure \ref{cosmodiag} while the second term is the pure kinematical correction term which is independent of the coupling function $\lambda(\tau)$.

\subsubsection{$\langle \gamma B_{\mu}B^{\mu}\rangle$}
\label{Sec_gBB}
In this section, we will calculate the correlator $\langle \gamma B_{\mu}B^{\mu}\rangle$. The covariant magnetic field $B_{\mu}$ is defined with respect to a comoving observer as in (\ref{ebnudef0}). Then   $\gamma B_{\mu}B^{\mu}$ can be expressed in terms of gauge field tensor as follows
 \bea
 \label{gBB-gAA}
 \gamma B_{\mu}B^{\mu}= \frac{1}{2a^4} \gamma F_{ij}F_{ij}-\frac{1}{a^4} \gamma \gamma_{ij} F_{il}F_{jl} + {\cal O}(\gamma^3)\; . 
\eea
One can evaluate the correlator by substituting (\ref{gBB-gAA}) and (\ref{H_gAA}) into (\ref{exp1_in-in}). 
\bea
\label{pertb_gBB}
\langle\gamma({\bf k}_1,\tau_I) B_{\mu}({\bf k}_2,\tau_I)B^{\mu}({\bf k}_3,\tau_I)\rangle &=& \lle  0\rb \gamma({\bf k}_1,\tau_I) B_{\mu}({\bf k}_2,\tau_I)B^{\mu}({\bf k}_3,\tau_I) \lb  0 \rgr \nonumber \\ &+& i\int_{-\infty}^{\tau_I}d\tau \lle  0\rb[H_{\rm int}(\tau),\gamma({\bf k}_1,\tau_I) B_{\mu}({\bf k}_2,\tau_I)B^{\mu}({\bf k}_3,\tau_I)]\lb  0 \rgr
\eea
As discussed in the case of the $\gamma A_{\mu}A^{\mu}$ correlator, the first term in the RHS of (\ref{pertb_gBB}) leads to  a non zero contribution to the vacuum expectation value of $\gamma B_{\mu}B^{\mu}$. This kinematical correction term is calculated using (\ref{gBB-gAA}) to yield the following contribution  
\begin{align}
\lle  0\rb \gamma({\bf k}_1,\tau_I) B_{\mu}({\bf k}_2,\tau_I)B^{\mu}({\bf k}_3,\tau_I) \lb  0 \rgr = - \frac{1}{2} (2\pi)^3\delta^{(3)} (\vect{k}_1+\vect{k}_2+\vect{k}_3)\epsilon_{ij}\frac{k_{2i}k_{2j}}{\tilde{k}_2^{2}}P_\gamma(k_1,\tau_I) P_B(\tilde{k}_2,\tau_I) 
\end{align}
where ${\bf \tilde{k}}_2={\bf k}_2+\frac{1}{2}{\bf k}_1$. Once again, we notice that the vacuum expectation value of $\gamma B_{\mu}B^{\mu}$ contributes at the leading order to the non-Gaussian correlator.  However, we would like to emphasise that these kinematical correction terms were not explored in earlier works \cite{Jain:2012vm,Shiraishi:2012xt}.

The leading order contribution of the second term in the RHS of (\ref{pertb_gBB}) is calculated using  the cosmological diagrammatic rules \cite{Giddings:2010ui}. Then one would obtain the total contribution as, 
\bea 
\label{finalgBB1}
\langle\gamma({\bf k}_1,\tau_I) B_{\mu}({\bf k}_2,\tau_I)B^{\mu}({\bf k}_3,\tau_I)\rangle & =&  -(2\pi)^3 \delta^{(3)}({\bf k_1}+{\bf k_2}+{\bf k_3})\epsilon_{ij}  \frac{k_{2i}k_{2j}}{k_2^2} \nonumber \\ 
&\times & \Biggl[ \frac{k_2^2\mathcal{I}_1}{a^4}+\left\lbrace({\bf k_2} + {\bf k_3})^2  -({\bf k_2} \cdot {\bf k_3})\right\rbrace \frac{k_2^2\mathcal{I}_2}{a^4}  \Biggr. \nonumber \\ 
 & + & \Biggl. \frac{1}{2}  \left(\frac{k_2}{\tilde{k}_2}\right)^2 P_{\gamma}(k_1,\tau_I)P_B(\tilde{k}_2,\tau_I) \Biggr],
\eea
where $\mathcal{I}_{1}$ and $\mathcal{I}_{2}$ are the same integrals, as given in (\ref{I1}) and (\ref{I2}).
By inserting (\ref{Int1}) and (\ref{Int2}) into (\ref{finalgBB1}), the final result for the correlation function becomes
\bea
\label{final_gBB2}
\lle \gamma({\bf k}_1) B_{\mu}({\bf k}_2)B^{\mu}({\bf k}_3)\rgr &=& (2\pi)^3 \delta^{(3)}({\bf k_1}+{\bf k_2}+{\bf k_3})\epsilon_{ij} \frac{k_{2i}k_{2j}}{k_2^2}\nonumber \\  
&\times &\Biggl[{\cal M}_B(k_1,k_2,k_3,\tau_I){\cal W}_B(k_1,k_2,k_3)
 -\frac{1}{2}  \left(\frac{k_2}{\tilde{k}_2}\right)^2 P_{\gamma}(k_1)P_B(\tilde{k}_2)\Biggr],
\eea
with 
\bea
{\cal M}_B(k_1,k_2,k_3,\tau_I)&=&\frac{|\gamma^{(0)}_{k_1}(\tau_I)|^2 |A^{(0)}_{k_2}(\tau_I)| |A^{(0)}_{k_3}(\tau_I)|}{a^4}\ =\ \frac{1}{a^2}{\cal M}_A(k_1,k_2,k_3,\tau_I),\\
{\cal W}_B(k_1,k_2,k_3)&=&k_2^3 k_3~\tilde{\mathcal{I}}_n^{(1)} +\left[({\bf k_2} + {\bf k_3})^2  -({\bf k_2} \cdot {\bf k_3})\right] k_2^2\,\tilde{\mathcal{I}}_n^{(2)}.
\eea
These contributions borrow similar structure as derived earlier in the case of the correlator $\left< \zeta\, {\bf B}\cdot {\bf B}\right>$ in \cite{Jain:2012ga, Jain:2012vm,Shiraishi:2012xt}. Note the presence of the two non-trivial integrals which can be exactly calculated for various values of $n$. However, the second term in (\ref{final_gBB2}) is the pure kinematical correction term which was not pointed out in earlier works \cite{Jain:2012vm,Shiraishi:2012xt}.


\subsubsection{$\langle \gamma E_{\mu}E^{\mu}\rangle$}

We now turn to the calculation of the correlator $\langle \gamma E_{\mu}E^{\mu}\rangle$, i.e., non-Gaussian cross correlation of tensor mode with the electric fields which is a bit more subtle than the other correlators. 
Using equation (\ref{ebnudef0}), one can express the covariant electric field in terms of gauge fields as
\bea
\label{E^2}
 E_{\mu}({\bf x},\tau)=\frac{d}{d\tau}A_{\mu}({\bf x},\tau)+i[H_{\rm int},A_{\mu}],
 \eea
where the extra commutator $i[H_{\rm int},A_{\mu}]$ arises due to the fact that the operators are all implicitly understood to be in the interaction picture, as discussed in Appendix \ref{dynamical}. With this definition, the observable $ \gamma E_{\mu}E^{\mu}$ is expressed as follows 
\bea
\label{gEE_Kine/dyn}
 \gamma E_{\mu}E^{\mu}=\frac{1}{a^4} \left[\gamma \frac{dA_i}{d\tau}\frac{dA_i}{d\tau}+i \gamma \left(\frac{dA_i}{d\tau}[H_{\rm int},A_i]+[H_{\rm int},A_i]\frac{dA_i}{d\tau}\right)-\gamma \gamma_{ij} \frac{dA_i}{d\tau}\; \frac{dA_j}{d\tau}\right]+ {\cal O}(\gamma^3)~.
\eea
Again, using the master formula of the in-in formalism (\ref{exp1_in-in}), we find that
\bea
\label{pertb_gEE}
\langle\gamma({\bf k}_1,\tau_I) E_{\mu}({\bf k}_2,\tau_I)E^{\mu}({\bf k}_3,\tau_I)\rangle &=& \lle  0\rb \gamma({\bf k}_1,\tau_I) E_{\mu}({\bf k}_2,\tau_I)E^{\mu}({\bf k}_3,\tau_I) \lb  0 \rgr \nonumber \\ &+& i\int_{-\infty}^{\tau_I}d\tau \lle  0\rb[H_{\rm int}(\tau),\gamma({\bf k}_1,\tau_I) E_{\mu}({\bf k}_2,\tau_I)E^{\mu}({\bf k}_3,\tau_I)]\lb  0 \rgr
\eea
As discussed earlier, the first term in the RHS of the above equation produces a non zero contribution to the vacuum expectation value of $\gamma E_{\mu}E^{\mu}$ due to the second and the third term in the RHS of equation (\ref{gEE_Kine/dyn}). The contributions arising from the second  term can be referred as the \textit{dynamical correction term} and the third term as the kinematical correction term, respectively.  In the super-horizon limit,  they can be calculated as follows, 
\bea 
 \lle  0\rb \gamma({\bf k}_1,\tau_I) E_{\mu}({\bf k}_2,\tau_I)E^{\mu}({\bf k}_3,\tau_I) \lb  0 \rgr &=& (2\pi)^3\delta^{(3)} (\vect{k}_1+\vect{k}_2+\vect{k}_3)\epsilon_{ij}\frac{k_{2i}k_{2j}}{k_2^2}\nonumber \\ &\times&\Bigg[- \left(\frac{k_2^2+k_3^2+{\bf k_2}\cdot {\bf k_3}}{ k_3^2}\right)\left(|A'_{k_3}(\tau_I)|^2 +|A'_{k_2}(\tau_I)|^2 \right)|\gamma_{k_1}(\tau_I)|^2\nonumber \\ 
 &+& \frac{1}{2}\left(\frac{k_2}{\tilde{k}_2}\right)^2 P_\gamma(k_1,\tau_I) P_E(\tilde{k}_2,\tau_I) \Bigg].
\eea
Unlike previous sections, the above vacuum expectation value is not only contributed by a kinematical correction term but also a dynamical correction term. Finally, the leading order contribution arising from the second term in the RHS of (\ref{pertb_gEE}) is calculated using the cosmological diagrammatic rules \cite{Giddings:2010ui} and the final result becomes  
\bea
\label{final_gEE2}
\lle \gamma({\bf k}_1) E_{\mu}({\bf k}_2)E^{\mu}({\bf k}_3)\rgr &=& (2\pi)^3 \delta^{(3)}({\bf k_1}+{\bf k_2}+{\bf k_3})\epsilon_{ij}\frac{k_{2i}k_{2j}}{k_2^2}\Bigg[ -\left(\frac{({\bf k_2} + {\bf k_3})^2-({\bf k_2} \cdot {\bf k_3})}{k_3^2}\right)\frac{\mathcal{J}_1}{a^4}-k_2^2 \, \frac{\mathcal{J}_2}{a^4} \nonumber \\
&-&\left(\frac{k_2^2+k_3^2+{\bf k_2}\cdot {\bf k_3}}{ k_3^2}\right)\left(\frac{|A'_{k_3}(\tau_I)|^2 +|A'_{k_2}(\tau_I)|^2 }{a^4}\right)|\gamma_{k_1}(\tau_I)|^2\nonumber \\ 
&+& \frac{1}{2}\left(\frac{k_2}{\tilde{k}_2}\right)^2 P_\gamma(k_1) P_E(\tilde{k}_2) \Bigg],
\eea
wherein the two integrals $\mathcal{J}_1$ and $\mathcal{J}_2$ are defined in terms of the mode functions of tensor mode and gauge field and their time derivatives as
\bea
\mathcal{J}_1&=& 2\,{\rm Im} \left[\gamma_{k_1}(\tau_I)A'_{k_2}(\tau_I)A'_{k_3}(\tau_I)\int d\tau  \lambda(\tau) \gamma_{k_1}^*(\tau) A_{k_2}^{'*}(\tau)A_{k_3}^{'*}(\tau)\right],\label{I1E} \\
\mathcal{J}_2 &=& 2\,{\rm Im} \left[\gamma_{k_1}(\tau_I)A'_{k_2}(\tau_I)A'_{k_3}(\tau_I)\int d\tau  \lambda(\tau) \gamma_{k_1}^*(\tau)A_{k_2}^*(\tau)A_{k_3}^*(\tau)\right].\label{I2E}
\eea
Again, it becomes useful to define the two auxiliary integrals $\tilde{\mathcal{J}}_n^{(1)}$ and $\tilde{\mathcal{J}}_n^{(2)}$ associated with $\mathcal{J}_1$ and $\mathcal{J}_2$, in a similar manner as earlier, by
\bea
\mathcal{J}_1&=&  -|\gamma^{(0)}_{k_1}(\tau_I)|^2 |A'^{(0)}_{k_2}(\tau_I)| |A'^{(0)}_{k_3}(\tau_I)| ~k_2 k_3~\tilde{\mathcal{J}}_n^{(1)} \label{IntJ1} \\
\mathcal{J}_2&=&  -|\gamma^{(0)}_{k_1}(\tau_I)|^2 |A'^{(0)}_{k_2}(\tau_I)| |A'^{(0)}_{k_3}(\tau_I)| ~ \tilde{\mathcal{J}}_n^{(2)}\label{IntJ2}
\eea
where $\gamma^{(0)}_{k}(\tau_I)$ and $A'^{(0)}_{k}(\tau_I)$ are the asymptotic super-horizon values of the mode functions of the tensor mode and the time derivative of mode functions of the gauge field respectively. In the limit corresponding to the end of inflation $|k \tau_I| \to 0$, we find that $\tilde{\mathcal{J}}_n^{(1)} \simeq \tilde{\mathcal{I}}_n^{(1)}$ and $\tilde{\mathcal{J}}_n^{(2)} \simeq \tilde{\mathcal{I}}_n^{(2)}$, as shown in appendix \ref{integrals}.  Note that, a few more terms appear in the result for the correlator $\langle \gamma E_{\mu}E^{\mu}\rangle$ as compared to $\langle \gamma B_{\mu}B^{\mu}\rangle$ since the covariant electric field is defined in terms of the time derivative of the gauge field while the covariant magnetic field only contains spatial derivatives of the gauge field. The final result can now be written in a compact form as 
\bea
\lle \gamma({\bf k}_1) E_{\mu}({\bf k}_2)E^{\mu}({\bf k}_3) \rgr &=& (2\pi)^3 \delta^{(3)}({\bf k_1}+{\bf k_2}+{\bf k_3})\epsilon_{ij}\frac{k_{2i}k_{2j}}{k_2^2}\nonumber \\ &\times&\Bigg[\sum_{\alpha=1}^{2} {\cal M}_{\alpha E}{\cal W}_{\alpha E}
+ \frac{1}{2}\left(\frac{k_2}{\tilde{k}_2}\right)^2 P_\gamma(k_1) P_E(\tilde{k}_2) \Bigg],
\label{final_gEE_compact}
\eea
where 
\bea
{\cal M}_{1E}(k_1,k_2,k_3,\tau_I)&=&\frac{|\gamma^{(0)}_{k_1}(\tau_I)|^2 |A'^{(0)}_{k_2}(\tau_I)| |A'^{(0)}_{k_3}(\tau_I)|}{a^4},\\
{\cal M}_{2E}(k_1,k_2,k_3,\tau_I)&=&-|\gamma_{k_1}(\tau_I)|^2\left( \frac{|A'_{k_2}(\tau_I)|^2 +|A'_{k_3}(\tau_I)|^2 }{a^4}\right),\\
{\cal W}_{1E}(k_1,k_2,k_3)&=&\left[({\bf k_2}+{\bf k_3})^2-({\bf k_2} \cdot {\bf k_3})\right]\frac{k_2 \tilde{\mathcal{I}}_n^{(1)}}{k_3} +k_2^2 \tilde{\mathcal{I}}_n^{(2)},\\
{\cal W}_{2E}(k_1,k_2,k_3)&=& \frac{k_2^2+k_3^2+{\bf k_2}\cdot {\bf k_3}}{ k_3^2}.
\eea
Note that, the above result has been obtained in the super-horizon limit. This correlator has a similar structure as $\left<\gamma B_{\mu}B^{\mu} \right>$, albeit an extra contribution which arises precisely due to the definition of the covariant electric field in equation (\ref{E^2}). We refer to this contribution as the dynamical correction term to the correlator. Moreover, the last term in equation (\ref{final_gEE_compact}) is similar to the kinematical correction term as appeared earlier for $\left<\gamma B_{\mu}B^{\mu} \right>$ in equation (\ref{final_gBB2}). As a result, the associated non-linearity parameter becomes constant in the squeezed limit, as we shall discuss in the following section. 


\subsection{The magnetic and electric non-linearity parameters}
\label{nl-param}

Following the  approach of our previous papers \cite{Jain:2012ga, Jain:2012vm} for the case of a cross-correlation of the comoving curvature perturbation with primordial gauge fields, we can similarly define convenient and useful non-linearity parameters to characterise the cross-correlations of the inflationary tensor perturbation with primordial gauge fields. Let us first define the bispectra  associated with $\left<\gamma B_{\mu}B^{\mu} \right>$ and $\left<\gamma E_{\mu}E^{\mu} \right>$ as,
\bea \label{bispectraB}
\left<\gamma(\vect{k}_1)B_{\mu}(\vect{k}_2)  B^{\mu}(\vect{k}_3)\right> &\equiv& (2\pi)^3\delta^{(3)}(\vect{k}_1+\vect{k}_2+\vect{k}_3)\,{\cal B}_{\gamma B B}(k_1,k_2, k_3)~,\\ \label{bispectraE}
\left<\gamma(\vect{k}_1)E_{\mu}(\vect{k}_2)  E^{\mu}(\vect{k}_3)\right> &\equiv&(2\pi)^3\delta^{(3)}(\vect{k}_1+\vect{k}_2+\vect{k}_3)\,{\cal B}_{\gamma E E}(k_1,k_2, k_3)~,
\eea
where the bispectra only depend on the magnitude of the wavevectors due to spatial isotropy of the background expansion. 
The strength of the magnetic and electric field bispectra can be characterised by defining the non-linearity parameters $b_{NL}^{\gamma}$ and $e_{NL}^{\gamma}$ as follows
\bea
\label{b_NLdef}
 {\cal B}_{\gamma B B}(k_1,k_2, k_3)&=&\frac{1}{2} b_{NL}^{\gamma}P_\gamma(k_1)\big[ P_B(k_2)+P_B(k_3)\big],\\
 \label{e_NLdef}
{\cal B}_{\gamma E E}(k_1,k_2, k_3)&=&\frac{1}{2} e_{NL}^{\gamma}P_\gamma(k_1)\big[ P_E(k_2)+P_E(k_3)\big],
\eea
where $P_\gamma$, $P_B$ and $P_E$ are the power spectra of the tensor perturbations (\ref{tensor_Pwrspctrm}), magnetic field (\ref{P_B}) and electric field (\ref{P_E}), respectively and the RHS of these definitions are written in a symmetric manner. 
If the two non-linearity parameters $b_{NL}^{\gamma}$ and $e_{NL}^{\gamma}$ are momentum independent, they correspond to a {\it local} shape of the bispectra in the Fourier space which can be obtained from the relations \cite{Jain:2012ga,Jain:2012vm}
\bea
\label{localExp_B}
{\bf B} &=& {\bf B}^{G} +\frac{1}{2}b_{NL}^{\gamma} \gamma^{G}  {\bf B}^{G}~,\\
\label{localExp_E}
{\bf E} &=& {\bf E}^{G} +\frac{1}{2}e_{NL}^{\gamma} \gamma^{G}  {\bf E}^{G}~,
\eea
where ${\bf B}^{G}$, ${\bf E}^{G}$ and $\gamma^{G}$ are the Gaussian random fields. In the absence of any parity violating interactions, as is the case in our scenario, the Gaussian tensor mode $\gamma^{G}$ represents both the polarization degrees of freedom associated with the graviton. Similar definitions also allow one to compute the higher order cross correlations of curvature perturbations with gauge fields \cite{Nurmi:2013gpa}.

\begin{figure}[!t] 
\begin{center}
\includegraphics[width=8cm, height=7cm]{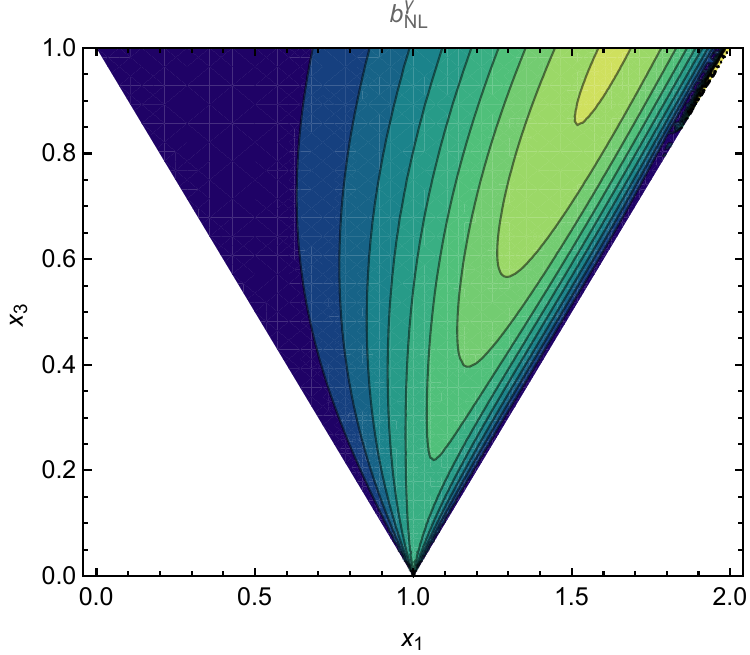}
\hskip 14pt
\includegraphics[width=8cm, height=7cm]{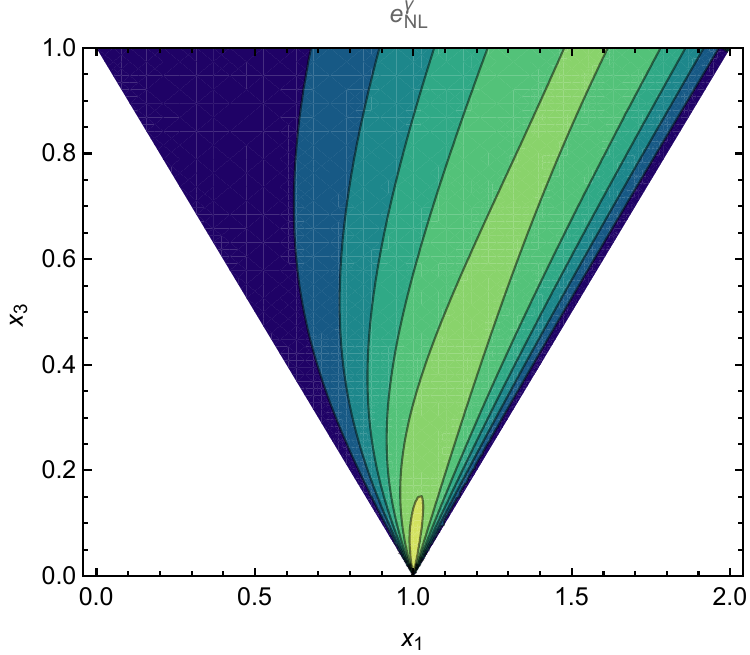}
\vskip 8pt
\begin{subfigure}{0.82\textwidth}
\includegraphics[width=6cm, height=1cm]{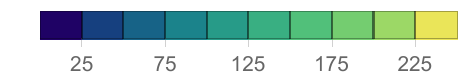}
\hskip 74pt
\includegraphics[width=6cm, height=1cm]{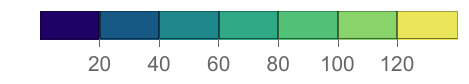}
\end{subfigure}
\end{center}
\caption{The extent of the non-linearity parameters $b_{NL} ^{\gamma}$ (left) and $e_{NL} ^{\gamma}$ (right) corresponding to different triangular configuration are plotted for the case of $n=2$. The colour legends representing the magnitude are also shown below each plot. While the top left corner of these plots corresponds to the squeezed limit, i.e., $(x_1, x_3)=(0,1)$, the top right corner represents the flattened shape, i.e., $(x_1, x_3)=(2,1)$. In the squeezed limit, $b_{NL} ^{\gamma}$ is relatively small while it is large in the nearly flattened shape. However, $e_{NL} ^{\gamma}$ remains small both in the exact squeezed and flattened limit but it becomes large as $x_3$ becomes smaller.} 
\label{bnl-enl}
\end{figure}

For the case of the $\langle \gamma B_{\mu}B^{\mu}\rangle$ correlator, we can now simply compare the equations (\ref{bispectraB}) and (\ref{b_NLdef}) with (\ref{final_gBB2}) and read off the bispectrum which leads to the following result for the non-linearity parameter 
$b_{NL}^{\gamma}$ for $n+1/2 >0$ as  
\bea
\label{b-nl}
b_{NL}^{\gamma} =\epsilon_{ij} \frac{k_{2i}k_{2j}}{k_2^2}  \Biggl[\left(\frac{(k_2 k_3)^{-(n+\frac{1}{2})}}{k_2^{1-2n}+k_3^{1-2n}}\right) {\cal W}_B(k_1,k_2,k_3) - \left(\frac{k_2}{\tilde{k}_2}\right)^2 \left(\frac{\tilde{k}_2^{{1-2n}}}{k_2^{1-2n}+k_3^{1-2n}}\right)\Bigg] ,
\eea
where, once again, the first term in the above result is due to the interaction Hamiltonian and the second term is the pure kinematical correction term.   
Similarly, on comparing (\ref{bispectraE}) and (\ref{e_NLdef}) with (\ref{final_gEE2}), yields the result for the non-linearity parameter 
$e_{NL}^{\gamma}$ for $n-1/2 >0$ as 
\bea
\label{e-nl}
e_{NL} ^{\gamma} =\epsilon_{ij} \frac{k_{2i}k_{2j}}{k_2^2}  \Biggl[\left(\frac{(k_2 k_3)^{(\frac{3}{2}-n)}}{k_2^{3-2n}+k_3^{3-2n}}\right) {\cal W}_{1E}(k_1,k_2,k_3)-{\cal W}_{2E}(k_1,k_2,k_3)+\left(\frac{k_2}{\tilde{k}_2}\right)^2 \!\! \left(\frac{\tilde{k}_2^{{3-2n}}}{k_2^{3-2n}+k_3^{3-2n}}\right)\Bigg].
\eea
These general expressions for $b_{NL} ^{\gamma}$ and $e_{NL} ^{\gamma}$ are one of the main results of this paper and they depend only on the wave numbers $k_1$, $k_2$ and $k_3$. 
The presence of $\delta^{(3)}(\vect{k}_1+\vect{k}_2+\vect{k}_3)$ in various cross-correlations constrains the wave vectors to form a closed triangle in the Fourier space. In order to understand the extent of these non-linearities, it is useful to study $b_{NL} ^{\gamma}$ and $e_{NL} ^{\gamma}$ for different shapes of the  triangle configuration. Conventionally, such shape functions are studied by  defining the dimensionless variables $x_1=\frac{k_1}{k_2}$ and $x_3=\frac{k_3}{k_2}$ while $k_2$ is set at an arbitrary scale. For an isotropic background, the shape function only depends  on any of the two momentum ratios. This allows us to first calculate the following two expressions as
\bea
\left(\frac{\tilde{k}_2}{k_2}\right)^2 &=& \frac{1}{2}\left(1+x_3^2-\frac{x_1^2}{2}\right),\\
\epsilon_{ij} \frac{k_{2i}k_{2j}}{k_2^2} &=& \frac{1}{\sqrt{2}}\left(\frac{(x_3^2-x_1^2-1)^2}{4x_1^2}-1\right).
\label{prefactor}
\eea
In order to calculate $b_{NL}^{\gamma}$ and $e_{NL}^{\gamma}$, we also need to evaluate the integrals $\tilde{\mathcal{I}}_n^{(1)}$ and $\tilde{\mathcal{I}}_n^{(2)}$ for different values of $n$ which has already been done in \cite{Jain:2012vm}. For the most interesting case of $n=2$ which leads to a scale invariant spectrum of magnetic fields, we present the results for $\tilde{\mathcal{I}}_n^{(1)}$ and $\tilde{\mathcal{I}}_n^{(2)}$ in Appendix \ref{integrals} and use them to plot the extent of these non-linearity parameters as contour plots in the momentum space in figure \ref{bnl-enl}. There are a few interesting points to note about the plots. First, both $b_{NL} ^{\gamma}$ and $e_{NL} ^{\gamma}$ remain very small in the squeezed limit which is characterised by the top left corner, i.e., $x_1 \to 0, x_3 \to 1$. Close to the flattened limit defined by $x_1 \to 2, x_3 \to 1$, $b_{NL} ^{\gamma}$ becomes large but $e_{NL} ^{\gamma}$ becomes very small in the exact flattened limit. However, $e_{NL} ^{\gamma}$ becomes large in a somewhat different limit $x_1 \to 1, x_3 \to 0$ in which $b_{NL} ^{\gamma}$ remains small.  Note that, the allowed region in the space of $x_1-x_3$ plane is constrained by the triangle inequality in the Fourier space. 

\subsection{The squeezed limit}

Let us now consider the squeezed limit wherein the wavelength of the graviton mode is much larger than the corresponding wavelengths of the electromagnetic fields. Thus, a squeezed shape non-Gaussianity implies a correlation of very large scale with very small scales. Using our general results of the in-in calculations from the previous sections, we can easily write down the results for all the correlators in the squeezed limit. However, as we shall discuss later, one can also obtain the squeezed limit non-Gaussianity by using semi-classical methods. In this limit, we have $\vect k_1\to 0$ and $\vect k_2 \to -\vect k_3 \equiv \vect k$. The two integrals of interest $\tilde{\mathcal{I}}_n^{(1)}$ and $\tilde{\mathcal{I}}_n^{(2)}$ simplify as 
\bea
\tilde{\mathcal{I}}_n^{(1)}&=& \begin{cases} \pi \int^{\tau_I} d\tau \tau J_{n-1/2}(-k\tau)Y_{n-1/2}(-k\tau), \quad n > -\frac{1}{2}\\ -\pi \int^{\tau_I} d\tau \tau J_{n-1/2}(-k\tau)Y_{n-1/2}(-k\tau), \quad n < -\frac{1}{2} \end{cases}\\
\tilde{\mathcal{I}}_n^{(2)}&=& \begin{cases}\pi \int^{\tau_I} d\tau \tau J_{n+1/2}(-k\tau)Y_{n+1/2}(-k\tau), \quad n > -\frac{1}{2}\\ -\pi \int^{\tau_I} d\tau \tau J_{n+1/2}(-k\tau)Y_{n+1/2}(-k\tau),  \quad n < -\frac{1}{2} \end{cases}
\eea
where $J_\alpha$ and $Y_\alpha$ are the Bessel functions of the first and second kind, respectively.  
These integrals are straightforward to evaluate and for integer values of $n$, one can show the following analytical result as\footnote{Note that, this result also holds true for real non-integer values of $n$ which can be verified numerically.} \cite{Jain:2012vm}
\beq\label{simpleint}
\pi \int^{\tau_I} d\tau \tau J_{n}(-k\tau)Y_{n}(-k\tau)= \frac{n}{k^2}, 
\eeq
and evidently, $\tilde{\mathcal{I}}_n^{(2)} = \tilde{\mathcal{I}}_{n+1}^{(1)}$.  In the squeezed limit, $\vect k_1 \to 0$, so using this result in all the correlators that we have obtained earlier, and after some simplification, we find the following results 
\bea 
\label{sqzd_gAA}
\langle \gamma(\vect{k}_1)A_{\mu}(\vect{k}_2)A^{\mu}(\vect{k}_3)\rangle &=&  \begin{cases}(2\pi)^3 \delta^{(3)}({\bf k_1}+{\bf k_2}+{\bf k_3})\epsilon_{ij}  \frac{k_{i}k_{j}}{k^2} \left(n+\frac{1}{2}\right)  P_{\gamma}(k_1)P_A(k)  ,\; \text{if} \; n > -\frac{1}{2}\\
-(2\pi)^3 \delta^{(3)}({\bf k_1}+{\bf k_2}+{\bf k_3})\epsilon_{ij}  \frac{k_{i}k_{j}}{k^2} \left(n-\frac{1}{2}\right)  P_{\gamma}(k_1)P_A(k), \;\text{if}\; n < -\frac{1}{2}\end{cases}
\eea
\bea 
\label{sqzd_gBB}
\langle \gamma(\vect{k}_1)B_{\mu}(\vect{k}_2)B^{\mu}(\vect{k}_3)\rangle &=&  \begin{cases}(2\pi)^3 \delta^{(3)}({\bf k_1}+{\bf k_2}+{\bf k_3})\epsilon_{ij}  \frac{k_{i}k_{j}}{k^2} \left(n-\frac{1}{2}\right)  P_{\gamma}(k_1)P_B(k)  ,\; \text{if} \; n > -\frac{1}{2}\\
-(2\pi)^3 \delta^{(3)}({\bf k_1}+{\bf k_2}+{\bf k_3})\epsilon_{ij}  \frac{k_{i}k_{j}}{k^2} \left(n+\frac{1}{2}\right)  P_{\gamma}(k_1)P_B(k), \;\text{if}\; n < -\frac{1}{2}\end{cases}
\eea
\bea 
\label{sqzd_gEE1}
\langle \gamma(\vect{k}_1)E_{\mu}(\vect{k}_2)E^{\mu}(\vect{k}_3)\rangle &=&  \begin{cases}(2\pi)^3 \delta^{(3)}({\bf k_1}+{\bf k_2}+{\bf k_3})\epsilon_{ij}  \frac{k_{i}k_{j}}{k^2} \left(n-\frac{1}{2}\right)  P_{\gamma}(k_1)P_E(k)  ,\; \text{if} \; n > \frac{1}{2}\\
-(2\pi)^3 \delta^{(3)}({\bf k_1}+{\bf k_2}+{\bf k_3})\epsilon_{ij}  \frac{k_{i}k_{j}}{k^2} \left(n+\frac{1}{2}\right)  P_{\gamma}(k_1)P_E(k), \;\text{if}\; n < \frac{1}{2}\end{cases}
\eea
The above results comprise of a set of novel consistency relations for the cross correlations involving the tensor perturbation and gauge fields. 
Note that, in the absence of other source terms, there exists an electromagnetic duality which states that the equations of motion and the power spectra of the electric and magnetic fields can be obtained from each other under a simultaneous exchange of ${\bf B} \to {\bf E}$, ${\bf E} \to -{\bf B}$ and $n \to -n$ \cite{Giovannini:2009xa}. To our knowledge, it has not been known if this duality is also maintained at the level of higher order correlation functions beyond the power spectra. In our case, we observe that this duality is indeed preserved in the squeezed limit of the cross correlations involving the electric and magnetic fields. In particular,  the results in equation (\ref{sqzd_gBB}) can be obtained from equation (\ref{sqzd_gEE1}) using the electromagnetic duality and vice versa. 
The results in both the regimes of the $\langle \gamma E_{\mu}E^{\mu}\rangle$ correlator can be correctly obtained from the corresponding results for the $\langle \gamma B_{\mu}B^{\mu}\rangle$ correlator under the electromagnetic duality transformations. In any case, it will be interesting to examine whether the electromagnetic duality is maintained for the higher order correlation functions away from the squeezed limit.

\section{Semi-classical derivation of the consistency relations } 
\label{semi-classical}

In this section, we want to apply semi-classical techniques to derive the three-point correlation function of the form $ \left<\gamma(\tau_I,\vect{k}_1) Y_{\mu}(\tau_I,\vect{k}_2) Y^{\mu}(\tau_I,\vect{k}_3)\right>$ for $Y_{\mu} = \left\{A_{\mu}, B_{\mu},  E_{\mu}\right\}$ in the squeezed limit $k_1 \ll k_2 \approx k_3$.
We would use the similar approach developed earlier to compute loop corrections to the primordial power spectrum during a de-Sitter/slow roll background \cite{Giddings:2010nc, Giddings:2011zd}. 
To start with, the correlator $\langle \gamma Y_{\mu}Y^{\mu}\rangle$ in real space can be written as  
\begin{align}
\left<\gamma(\vect x)Y_{\mu}({\vect x})Y^{\mu}({\vect x})\right>= \int \frac{ d^3 \vect k_1}{(2\pi)^3} \int \frac{ d^3 \vect k_2}{(2\pi)^3}\int    \frac{d^3 \vect k_3}{(2\pi)^3}  e^{i \vect x\cdot (\vect k_1+\vect k_2+\vect k_3)}  \left<\gamma(\vect k_1)Y_{\mu}(\vect k_2)Y^{\mu}(\vect k_3)\right>.
\end{align}
 In the limit where one of the momenta is much smaller than the other two i.e. $k_1 \ll k_2 \approx k_3$, the wavelength corresponding to the graviton $\gamma$ is very large and is frozen to a constant amplitude $\gamma^B$ outside the horizon. As a result, the effect due to a long wavelength graviton mode $\gamma^B$ on the two-point correlator of gauge fields can be understood by  rescaling the background as 
$
ds^2 =-dt^2+a^2(t) d \tilde{x}^2
$
with $ d\tilde x^2 \to  dx^2+\gamma^B_{ij}dx^idx^j$. 
Thus we can compute the two point correlation function $\langle Y_{\mu}Y^{\mu}\rangle$ in the modified background using the similar arguments of \cite{Giddings:2010nc}, by Taylor expanding around the unperturbed background in real space as
 \bea 
 \label{Taylorexpansion}
\left<Y_{\mu}(\vect x)Y^{\mu}(\vect x)\right>_B = \left<Y_{\mu}(\vect x)Y^{\mu}(\vect x)\right>_0+\gamma_{ij}^B\frac{\p}{\p \gamma^B_{ij}}\left.\left<Y_{\mu}({\tilde {\bf x}})Y^{\mu}({\tilde {\bf x}})\right>\right|_{\gamma^B=0}+\dots\,,
\eea
and in the Fourier space  as
\bea \label{qwas}
\left<Y_{\mu}(\vect x)Y^{\mu}(\vect x)\right>_B &=& \int\int \frac{ d^3 \vect k_2}{(2\pi)^3}  \frac{d^3 \vect k_3}{(2\pi)^3}   e^{i \vect x\cdot\vect k_2} e^{i \vect x\cdot\vect k_3}\left<Y_{\mu}({\vect k_2})Y^{\mu}({\vect k_3})\right>_B\nonumber\\
&=& \left<Y_{\mu}(\vect x)Y^{\mu}(\vect x)\right>_0 \nonumber\\
& &+ \left(\gamma^B_{ij}(\vect x_0)\frac{\p}{\p\gamma^B_{ij}}+\frac{1}{2}\gamma^B_{ij}(\vect x_0)\gamma^B_{lk}(\vect x_0)\frac{\p}{\p\gamma^B_{ij}}\frac{\p}{\p\gamma^B_{kl}} \right)\nonumber\\
& & \times\left.\left[ \int\int \frac{d^3\tilde{\vect q}_1}{(2\pi)^3}\frac{d^3 \tilde {\vect q}_2}{(2\pi)^3}e^{i\vect{\tilde x} \cdot\vect{ \tilde  q}_1}e^{i\tilde{\vect x} \cdot \vect{\tilde q}_2}\left<Y_{\mu}({\tilde {\vect q}}_1)Y^{\mu}({\tilde {\vect q}}_2)\right>_B\right]\right|_{\gamma^B=0}\ .
\eea
Since the long wavelength mode $\gamma_B$ is almost constant over the Hubble horizon, and in particular, over the scales of variation of the short wavelength modes, we can choose   $\vect x_0 =\vect x $. 
We can now express the modified coordinates due to the long wavelength graviton as 
\beq
{\tilde x}^2 = \left[ e^{\gamma^B}\right]_{ij}x^i x^j =  x^2+\gamma^B_{ij}x^i x^j+\dots~,
\eeq
and using that momentum transforms inversely, define
\beq
{\tilde q}^2 = \left[ e^{-\gamma^B}\right]^{ij}q_i q_j =  q^2-\gamma^B_{ij}q_i q_j+\dots~,
\eeq
such that $\vect{\tilde x}\cdot \vect{\tilde q} = \vect x\cdot \vect q$, and
\beq
\frac{\p}{\p\gamma_{ij}^B}=\frac{\p\tilde q^2}{\p\gamma_{ij}^B}\frac{\p}{\p\tilde q^2} = -q_iq_j\frac{\p}{\p \tilde q^2}~.
\eeq
It is useful to note that $\det [\exp(\gamma^B)]=1$, so that the Jacobian of the transformation $\tilde q \to q$ is trivial, and $\delta^{(3)}(\tilde{\vect q}_1+\tilde{\vect q}_2) = \delta^{(3)}(\vect{q}_1+\vect{q}_2)$, so we can write
\bea
\label{tpf-mb}
\left<Y_{\mu}(\vect{\tilde q}_1)Y^{\mu}(\vect{\tilde q}_2)\right>_B
&=&(2\pi)^3\delta^{(3)}(\vect{q}_1+\vect{q}_2)\left[1+\left(-\gamma^B_{ij}q_iq_j +\frac{1}{2}\gamma^B_{il}\gamma^B_{lj}q_iq_j+\dots\right)\frac{\p}{\p q^2}\right.\nonumber\\
& &\left.+\frac{1}{2}
\left((\gamma^B_{ij}q_iq_j)^2+\dots\right) \frac{\p}{\p q^2}\frac{\p}{\p q^2}+\dots\right] P_Y(q).
\eea
In the squeezed limit, due to the rescaled background by the long wavelength graviton mode, one can write a three point correlation function in terms of the modified two point function as
\bea
\label{tpf-general}
\lim_{k_1 \to 0}\left<\gamma(\tau_I,\vect{k}_1) Y_{\mu}(\tau_I,\vect{k}_2) Y^{\mu}(\tau_I,\vect{k}_3)\right> = \left<\gamma(\tau_I,\vect{k}_1) \left<Y_{\mu}(\tau_I,\vect{k}_2) Y^{\mu}(\tau_I,\vect{k}_3)\right>_B\right>.
\eea
Let us now justify that the arguments used to obtain the consistency relations might only be trusted for specific regions of $n$. To illustrate this, let's write the evolution equation for the canonically normalized gauge field, i.e., equation (\ref{cngf}) again as
\beq\label{meq}
v_k'' +\left(k^2-\frac{S''}{S}\right)v_k=0~.
\eeq
This equation is identical to the evolution equation of a massless scalar field in the FLRW spacetime. However, the only difference is that the gauge field experiences an effective scale factor $S$ which we refer to as the {\it pump} field. For the massless scalar field in de-Sitter with $a \sim 1/\tau$, there exists an effective event horizon and therefore, the long and the short wavelength modes decouple on the super-horizon scales. This allows one to capture all the physical effects of the long wavelength mode by rescaling the background for the short wavelength modes to calculate the three point cross-correlations involving long and short wavelength modes of the scalar field fluctuations or the curvature perturbations. Similar arguments can be applied for any other correlator in the squeezed limit as far as the super-horizon modes are fixed by the values of the background quantities at horizon crossing.

If we inspect the mode function equation (\ref{meq}) above in the long wavelength limit $k^2 \ll S''/S$, the solution can be found by simple integration to be given by 
\beq
v_k(\tau) \sim C_1 S(\tau) + C_2 S(\tau) \int \frac{d\tau}{S^2(\tau)}
\eeq
where $C_1$ and $C_2$ are $k$-dependent integration constants. It is clear that if $S = {\sqrt{\lambda}} \sim \tau^{-n}$, then for $\tau\to 0$, the first term will dominate for $n > -1/2$, which means that the solution for $A_k = v_k/S$ will freeze to a constant in the long wavelength limit. A trivial rescaling of the field is of course not important, what is important is that the solution is in a super-horizon attractor regime. If the solution is dominated by the attractor, then the rescaling of the momentum by a long mode will enter trivially as a rescaling of the initial normalization through $C_1$. However, if the non-attractor is important, the rescaling of momentum will also enter through the lower limit of the time integral multiplying $C_2$ and thus modify the dynamics non-trivially. The existence of an attractor solution of the long wavelength modes leads to the constraint $n > -1/2$. Therefore, the semi-classical derivation for $\langle \gamma A_{\mu}A^{\mu}\rangle$ can only be trusted  for $n>-1/2$. A similar condition can also be obtained for the corresponding correlator involving the graviton and magnetic fields. However, we find that the properly rescaled electric field experiences an effective scale factor, $\frac{1}{\sqrt{\lambda}}$, which reverses the condition on $n$ for electric field. As a result, we conclude that the semi-classical derivation for the graviton electric fields cross-correlator can only be trusted for $n<1/2$.

By substituting equation (\ref{tpf-mb}) into equation (\ref{tpf-general}) and after further simplifications, we will obtain the desired correlators in the squeezed limit. For $n>-1/2$, we get the following consistency relations
   \beq
 \lim_{k_1 \to 0} \left<\gamma(\tau_I,\vect{k}_1) A_{\mu}(\tau_I,\vect{k}_2) A^{\mu}(\tau_I,\vect{k}_3)\right>
 = (2\pi)^3\delta^{(3)} (\vect{k}_1+\vect{k}_2+\vect{k}_3) 
 \left( n+\frac{1}{2}\right) \epsilon_{ij} \frac{k_{2i}k_{2j}}{k_2^2}P_\gamma(k_1) P_A(k_2),  \label{SC_A}
 \eeq
 \beq
 \lim_{k_1 \to 0} \left<\gamma(\tau_I,\vect{k}_1) B_{\mu}(\tau_I,\vect{k}_2) B^{\mu}(\tau_I,\vect{k}_3)\right>
 = (2\pi)^3\delta^{(3)} (\vect{k}_1+\vect{k}_2+\vect{k}_3) 
 \left( n-\frac{1}{2}\right) \epsilon_{ij} \frac{k_{2i}k_{2j}}{k_2^2}P_\gamma(k_1) P_B(k_2),  \label{SC_B}
 \eeq
 Similarly, for $n < 1/2$, we get 
  \beq
 \lim_{k_1 \to 0} \left<\gamma(\tau_I,\vect{k}_1) E_{\mu}(\tau_I,\vect{k}_2) E^{\mu}(\tau_I,\vect{k}_3)\right>
 = -(2\pi)^3\delta^{(3)} (\vect{k}_1+\vect{k}_2+\vect{k}_3) 
 \left( n+\frac{1}{2}\right) \epsilon_{ij} \frac{k_{2i}k_{2j}}{k_2^2}P_\gamma(k_1) P_E(k_2).  \label{SC_E}
 \eeq
In the squeezed limit, these consistency relations are quite general as they have been derived only by using the semi-classical techniques. These relations do exactly match with the squeezed limit results of the full in-in correlates, as obtained in (\ref{sqzd_gAA}), (\ref{sqzd_gBB}) and (\ref{sqzd_gEE1}).  From these correlators, one can now read off the non-linearity parameters  $b^{\gamma}_{NL}$ and  $e^{\gamma}_{NL}$ as
 \bea
 b^{\gamma}_{NL} &=&\left( n-\frac{1}{2}\right) \epsilon_{ij} \frac{k_{2i}k_{2j}}{k_2^2}, \quad n>-1/2 \\
 e^{\gamma}_{NL} &=&-\left( n+\frac{1}{2}\right) \epsilon_{ij} \frac{k_{2i}k_{2j}}{k_2^2}, \quad n<1/2. 
 \eea    
It is important to mention here that  these are indeed the {\it local} non-linearity parameters as the factor $\epsilon_{ij} k_{2i}k_{2j}/{k_2^2}$ is momentum independent in the squeezed limit, as evident from equation (\ref{prefactor}).  Moreover, both the parameters in the squeezed limit are proportional to $n$ and are of order unity as expected, following the discussion of our earlier paper \cite{Jain:2012ga}. This behaviour is also quite evident from the contour plots in figure \ref{bnl-enl} wherein the top left corner indicates the extent of these non-linearity parameters in the squeezed limit.

\section{A novel correlation of tensor and curvature perturbations}
\label{twocorr}

As mentioned in the introduction, an interesting consequence of these three point correlators involving the primordial tensor mode and gauge fields is the possibility of a novel two point correlation of the primordial tensor mode with the primordial curvature perturbation. As we shall show later, such a two point correlation function can actually be non-vanishing in the anisotropic background created by the long wavelength gauge fields, and may even receive contributions of the form $\left<\gamma {\bf B\cdot B}\right>$ from higher order quantum gravity corrections.

The presence of primordial gauge fields can leave indirect imprints on the spectrum of curvature perturbation during inflation. This can be illustrated by splitting the total curvature perturbations into the ``usual” contribution, $\zeta_0$ generated independently of the gauge field, and the contributions imprinted by magnetic and electric fields,
\bea
\zeta(\tau)=\zeta_0(\tau)+\zeta_B(\tau)+\zeta_E(\tau),
\eea
where $\zeta_B$ and $\zeta_E$ are  induced by the corresponding magnetic and electric fields derived from the gauge fields only. Now, we can express the two-point correlation function of  the tensor mode with the curvature perturbation as follows,
\beq
\left< \gamma({\bf k_1})\zeta({\bf k_2})\right> = \left< \gamma({\bf k_1})\zeta_B({\bf k_2})\right>+ \left< \gamma({\bf k_1})\zeta_E({\bf k_2})\right>.
\eeq
To this end, we are going to generically denote  $\zeta_B$ and $\zeta_E$ by  $\zeta_Y$. Following \cite{Nurmi:2013gpa}, they can be written in the Fourier space as
\bea
\zeta_Y({\bf k},\tau_I) = \int_{\tau_0}^{\tau_I}d\ln\tau \lambda(\tau)\int\frac{d^3 \vect q}{(2\pi)^3}\frac{Y_i({\bf q})Y^i({\bf k-q})}{3H^2 \ep}~,
\eea
where $\epsilon$ is the first slow roll parameter, $H$ is the Hubble parameter, and $Y=\{E, B\}$.  We can now express the two-point correlator $\left<\gamma \zeta_Y\right>$ in terms of the three point function $\langle \gamma Y_{\mu}Y^{\mu}\rangle$ as
\bea
\label{gammazeta}
\left< \gamma({\bf k_1},\tau_I)\zeta_Y({\bf k_2},\tau_I)\right>  = \int_{\tau_0}^{\tau_I}d\ln\tau \lambda(\tau)\int\frac{d^3 \vect q}{(2\pi)^3}\frac{\left< \gamma({\bf k}_1,\tau_I)Y_i({\bf q},\tau)Y^i({\bf k_2-q},\tau)\right>}{3H^2\ep}~.
\eea
To compute $\left<\gamma \zeta\right>$, we need to evaluate this three dimensional integral in Fourier space. However, note that, these three point correlators appearing in the integrand of above equation are not equal time correlation functions \cite{Nurmi:2013gpa}. Thus, we first have to convert them in terms of an equal time correlation function as
\bea
\label{eq.time_B}
\left< \gamma({\bf k_1},\tau_I)\zeta_Y({\bf k_2},\tau_I)\right>  = \int_{\tau_0}^{\tau_{I}}d\tau \frac{\lambda(\tau_I)}{\tau}\left(\frac{\tau}{\tau_I}\right)^{\alpha_Y}\int\frac{d^3 \vect q}{(2\pi)^3}\frac{\left< \gamma({\bf k}_1,\tau_I)Y_i({\bf q},\tau_I)Y^i({\bf k_2-q},\tau_I)\right>}{3H^2\ep} \nonumber \\ \times \; \Theta(|{\bf k_2}-{\bf q}|-k_0)\Theta(q-k_0)\Theta\left(-\frac{1}{\tau}- q\right)\Theta\left(-\frac{1}{\tau}- |{\bf k_2}-{\bf q}|\right), 
\eea
where $\Theta$ is the Heaviside step function and 
\bea
\alpha_B&=&\begin{cases} 4-2n, \; n>-1/2 \\
6+2n, \; n<-1/2
\end{cases}\\
\alpha_E &=&\begin{cases} 6-2n, \; n>1/2 \\
4+2n, \; n<1/2
\end{cases}
\eea
Here, we have assumed that only super-horizon modes of gauge fields are  contributing non trivially to these integrals with  $\theta$ functions stressing this fact and $k_0 \equiv a_0 H_0$ is the horizon crossing scale at the time $\tau_0$ which we assume to be the onset of the generation of magnetic fields. 
In general, the momentum integral appearing in equation (\ref{eq.time_B}) is somewhat non-trivial to evaluate. However,  without loss of generality, one can analytically calculate these two-point correlators by first expressing these three-point correlators as   
\bea 
\lle\gamma^{(s)}({\bf k}_1)Y_i({\bf k}_2)Y^i({\bf k}_3)\rgr=(2\pi)^3 \delta^{(3)} ({\bf k}_1+{\bf k}_2+{\bf k}_3)\ep^{s}_{ij}(\hat{k}_1)\hat{k}_{2i}\hat{k}_{2j} F(k_1, k_2, k_3),
\eea
and using it in equation  (\ref{eq.time_B}) to arrive at
\bea
\left< \gamma({\bf k_1},\tau_I)\zeta_Y({\bf k_2},\tau_I)\right>  =\delta^{(3)} ({\bf k}_1+{\bf k}_2) \int_{\tau_0}^{\tau_{I}}d\tau \frac{\lambda(\tau_I)}{\tau}\left(\frac{\tau}{\tau_I}\right)^{\alpha_Y}\int d^3 \vect q\; \ep_{ij}^{s} \hat{q}_i \hat{q}_j \frac{F(k_1,q, |{\bf k}_2-{\bf q}|)}{3H^2\ep} \nonumber \\ \times \; \Theta(|{\bf k_2}-{\bf q}|-k_0)\Theta(q-k_0)\Theta\left(-\frac{1}{\tau}- q\right)\Theta\left(-\frac{1}{\tau}- |{\bf k_2}-{\bf q}|\right).
\eea
One can now choose a convenient coordinate system to evaluate the momentum integral. We choose a spherical polar coordinate system in such a way that its polar axis is along the direction of ${\bf k}_2$. 
This allows us to construct a triplet of orthonormal basis vectors $\{ \hat{{\bf e}}_1, \hat{{\bf e}}_2, \hat{{\bf k}}_2 \}$ associated with $\epsilon^{(s)}_{ij}(\hat {{\bf k}}_2)$. Thus, any arbitrary unit vector  $\hat{{\bf q}}$  in such a spherical polar coordinate system can be expressed as follows
\bea 
\label{unit_vect}
\hat{{\bf q}}=\sin \theta \cos \phi \; \hat{{\bf e}}_1+\sin \theta \sin \phi \; \hat{{\bf e}}_2+\cos \theta \; \hat{{\bf k}}_2~.
\eea
Using this expression and through a straightforward calculation, we can now obtain the following relation  
\bea 
\ep_{ij}^{s} \hat{q}_i \hat{q}_j=\frac{1}{\sqrt{2}} \sin^2 \theta \, e^{i s \phi}~.
\eea
Since $F$ is independent of the angle $\phi$ and so are all the $\Theta$ functions, we can evaluate the $q$-integral and obtain (for brevity, we are not writing all the $\Theta$ function terms again)
\bea 
\int d^3 \vect q\; \ep_{ij}^{s} \hat{q}_i \hat{q}_j \frac{F(k_1,q, |{\bf k}_2-{\bf q}|)}{3H^2\ep}= \frac{1}{3 \sqrt{2}}\int_{0}^{2\pi} d\phi \; e^{i s \phi} \int_0^{\infty} dq\; q^2 \int_0^{\pi} d\theta \sin^3 \theta  \frac{F(k,q, |{\bf k}-{\bf q}|)}{H^2\ep}.
\eea
Interestingly, this integral identically vanishes because of the $\phi$ integral and as a result, the direct two-point correlation of the tensor mode with the induced curvature perturbation also vanishes i.e. 
\bea
\left< \gamma({\bf k_1})\zeta({\bf k_2})\right>=\left< \gamma({\bf k_1})\zeta_Y({\bf k_2})\right>=0.
\eea
Furthermore, one can also show the above result by performing the angular integral using the identities involving the spin-weighted spherical harmonics and using the transverse and traceless condition of the polarization tensor, along the lines of the discussion presented in \cite{Shiraishi:2012xt}.

This does, however, not mean that we should necessarily expect the $\left<\gamma \zeta\right>$ correlator to vanish. It is worth noting that isotropy is broken by the long wavelength modes of the vector field. As long wavelength modes are stretched to super-horizon scales, they act as a background for the shorter scales. Thus, if inflation lasts for a total number of e-folds, $N_{\textrm{tot}}$, the modes which left the horizon during the first $(N_{\textrm{tot}}-N_*)$ e-folds, they contribute as a classical background for the modes leaving the horizon after $N_*$ e-folds remain. These long wavelength modes make up a random vector pointing in some random direction, which locally breaks isotropy.  The typical amplitude in a given local realization is given by the square root of the variance of the long wavelength fluctuations \cite{Kofman:1986wm,Bartolo:2012sd}. Let us, therefore, assume that in the case of inflationary magnetogenesis, a background magnetic field is turned on in the x-direction. This is not completely general, as we already chose our coordinate system to be aligned with $\bf{k}$, but it serves to show the point we want to make. In any case, we will take
\beq
B_i = B_c \delta_{ix} + \delta B_i ~,\qquad E_i = \delta E_i~.
\eeq
However, the discussion proceeds very similar for the case of the electric field being amplified on super-horizon scales. 
We will turn our attention to two interesting contributions to the $\left< \gamma({\bf k_1})\zeta({\bf k_2})\right>$ correlation function. This first comes from the induced direct coupling between $\gamma$ and $\zeta$, while the second is a contribution arising from a quantum gravity induced higher dimensional operator such that the effective action is written as 
\beq
S=-\frac{1}{4}\int d^4x \sqrt{-g}\, \lambda(\phi) \left(F_{\mu \nu}F^{\mu \nu}+\frac{1}{4M^4}(F_{\mu \nu}F^{\mu \nu})^2+\cdots\right),
\eeq
with $M$ being the scale of quantum gravity.  The first contribution comes from expanding the coupling $\lambda(\phi)$ in perturbations of the scalar field, using the comoving gauge \cite{Ferreira:2014zia}
\beq
\phi = \left.\phi\right|_{\delta\ln a =0} +\left.\frac{\partial \phi}{\partial \ln a}\right|_{\delta\ln a =0} \delta \ln a +\dots \ =\  \left.\phi\right|_{\zeta  =0} + \left.\frac{\partial \phi}{\partial \ln a}\right|_{\zeta =0} \zeta+\dots~,
\eeq
so we find the interaction Hamiltonian
\beq
H_{\textrm{int}}^{(1)} =  -\frac{1}{2}  {B_c^2} a\, \xi \int d^3 x\zeta \gamma_{xx},
\eeq
where we have defined
\beq
\xi (\tau) = \frac{\partial \lambda}{\partial \phi} \frac{\phi'(\tau)}{H}~.
\eeq
Using the cosmological diagrammatic rules \cite{Giddings:2010ui}, we arrive at 
\beq
\left< \gamma({\bf k}_1)\zeta({\bf k}_2)\right>= (2\pi)^3\delta^{(3)}({\bf k}_1+{\bf k}_2) \sqrt{2}\, \textrm{Im}\left[ \gamma_{k_1}(\tau_I)\zeta_{k_2}(\tau_I)\int d\tau a\, \xi B_c^2 \gamma^*_{k_1}(\tau)\zeta^*_{k_2}(\tau)\right],
\eeq
where we used $\epsilon_{xx} =1/\sqrt{2}$. This integral will not vanish in general, but its detailed value will depend on the time dependence of $\xi$ and $B_c$, as previously discussed in \cite{Chen:2014eua, Choi:2015wva}.

The second contribution coming from a higher order quantum gravity term of the form $\frac{1}{4 M^4}(F_{\mu\nu}F^{\mu\nu})^2$, leads to an interaction Hamiltonian
\beq
H_{\textrm{int}}^{(2)} = -\frac{1}{2 M^4}\int d^3 x \lambda(\tau)  \Big(\gamma_{ij}E_iE_j -\gamma_{ij}B_iB_j\Big)\left(E^2 -B^2\right),
\eeq
to linear order in $\gamma$. Expanding this to the quadratic order in the electromagnetic field perturbations, the anisotropic part of the interaction Hamiltonian reads 
\beq
H_{\textrm{int, aniso}}^{(2)} = \frac{1}{2 M^4}\int d^3 x \lambda(\tau)  B_c^2 \left[-\gamma_{xx}(\delta E^2-\delta B^2) +4 \gamma_{ix}\delta B_i \delta B_x \right].
\eeq
Apart from the changed polarization sum, this interaction Hamiltonian is similar to equation (\ref{H_gAA}), if we take the effective coupling as
\beq
\lambda \to \tilde \lambda = \frac{B_c^2}{a^2 M^4} \lambda.
\eeq
Upon using the same techniques as earlier, we now obtain
\beq
\label{gBB-aniso}
\left< \gamma({\bf k_1})B_{\mu}({\bf k_2})B^{\mu}({\bf k_3})\right>_{\text{aniso}} = (2\pi)^3\delta^{(3)}({\bf k}_1+{\bf k}_2+{\bf k}_3)\left[\epsilon_{xx} F^{(2)}(k_1, k_2, k_3) +\epsilon_{ix} G^{(2)}(k_1, k_2, k_3)\right],
\eeq
where 
\bea
F^{(2)}(k_1, k_2, k_3)  &=& 2({\bf k}_2 \cdot {\bf k}_3 )\frac{\mathcal{I}_1}{a^4}+\left[({\bf k}_2 \cdot {\bf k}_3)^2 +5 k_2^2k_3^2\right] \frac{\mathcal{I}_2}{a^4},\nonumber\\
 G^{(2)}(k_1, k_2, k_3)  &=& \left[2({\bf k}_2 \cdot {\bf k}_3)(k_{2x}k_{3i}+k_{3x}k_{2i})-4(k_{3i}k_{3x}k_2^2+k_{2i} k_{2x}k_3^2)\right] \frac{\mathcal{I}_2}{a^4},
\eea
and the integrals $\mathcal{I}_1$ and $\mathcal{I}_2$ are defined in the same way as earlier but with the effective coupling $\tilde \lambda$ replacing $\lambda$.
Note that, the structure of this result is quite similar to the previous result in equation (\ref{final_gBB2}), however, there are crucial differences.  Since the background magnetic field is turned on in the x-direction, upon using equation (\ref{gBB-aniso}) in equation (\ref{eq.time_B}), the angular integrals  can be trivially carried out which will lead to a non-vanishing result for the $\left<\gamma \zeta\right>$ correlator. Therefore, such an imprint induced in this case can be considered as a novel consequence of such scenarios with higher order quantum gravity corrections.


\section{Conclusions and discussions}
\label{conclusions}  

Higher order cosmological correlations generally provide new insights into the non-trivial interactions in the early universe.
This paper has calculated a new set of such cosmological correlation functions and their corresponding semi-classical consistency relations.
More precisely, we have computed the non-Gaussian correlation functions of primordial gravitons produced during inflation, with gauge fields that are non-minimally coupled in the early universe. Our correlation functions are derived, both in terms of the vector potential, corresponding to the correlation of gravitons with a vector boson, and in terms of the associated electric and magnetic fields for the $U(1)$ gauge boson. We have obtained the full general results for these cross-correlations for the first time in the literature by taking into account the correct time evolution operator in the interaction picture of the in-in formalism. 
Moreover, we have also shown the presence of the leading order correction terms in all the correlators that had not been noticed earlier. In the squeezed limit, we found that the non-linearity parameters remain small but can be  large for the other triangular shapes of the bispectra in Fourier space.

In order to check our results, we have derived new semi-classical relations analogous to known semi-classical relations in the literature (also sometimes called the consistency relations or soft theorems, etc.). Some well-known examples are the Maldacena consistency relations \cite{Maldacena:2002vr} for the three-point correlation functions, the SSV relations \cite{Seery:2008ax} for the four-point functions, and the GS relations for the infrared part of loop diagrams \cite{Giddings:2010nc}. These relations are generally valid for curvature perturbation modes and graviton modes, but they are not directly translatable to the present case. In the minimally coupled case, the gauge field sector is invariant under the conformal transformations and therefore does only feel the expansion of spacetime through trivial conformal rescaling of the flat spacetime results. 
Thus, it is evident that the non-trivial contribution to the correlation functions arises due to the time dependence of the non-minimal coupling that breaks the conformal invariance. Therefore all the correlation functions computed here are proportional to the time derivative of the non-minimal coupling function. However, a suitable redefinition of the {\it pump} field in terms of the non-minimal coupling function and a gauge field redefinition allows us to write the action for the polarization modes of the gauge fields in the identical form to a scalar field in a fictitious expanding universe, with an expansion given by the time dependence of the non-minimal coupling.
For a non-minimal coupling corresponding to a fictitious inflationary universe, we show that our new semi-classical relations correctly recover the squeezed limit of the full in-in calculations. We checked the agreement with our new consistency relations both for the vector field itself, and also in terms of the corresponding electric and magnetic fields.

In particular, it is worthwhile keeping in mind, that even if the gauge field is not directly observable today, as in the case of dark sector gauge fields, it may be indirectly observable through its imprints on the curvature perturbations.  Finally, we have calculated a direct correlation between one graviton mode and a curvature perturbation mode, induced by the three-point correlation function of one graviton mode and two gauge field modes, and showed that it vanishes in the isotropic limit.  However, as we briefly discussed in the previous section, this two-point correlator is actually non-vanishing in general due to the anisotropic background of long wavelength gauge fields and also receives non-trivial contributions in non-linear theories of electrodynamics arising in quantum gravity. 
It would further be interesting to study the observational consequences of such correlators in terms of non-zero cross-correlations involving temperature and polarization anisotropies in the CMB and understand the prospects of their detectability with both ground and space based upcoming CMB experiments.


\section*{Acknowledgments}

RKJ would like to acknowledge financial support from the new faculty seed start-up grant of IISc, the Core Research Grant CRG/2018/002200 from the Science and Engineering Research Board, Department of Science and Technology, Government of India and the Infosys Foundation, Bangalore through the Infosys Young Investigator award.  MSS is supported by Villum Fonden grant 13384 and Independent Research Fund Denmark grant 0135-00378B.


\appendix

\section{Useful integrals}
\label{integrals}
For $n>-\frac{1}{2}$, the integrals $\tilde{\mathcal{I}}_n^{(1)}$ and $\tilde{\mathcal{I}}_n^{(2)}$ can be explicitly written as, 
\bea
\tilde{\mathcal{I}}_n^{(1)}&=& \frac{\pi^3}{2}\frac{2^{-2n-1}}{\Gamma^2(n+1/2)}(-k_2\tau_I)^{n+1/2}(-k_3\tau_I)^{n+1/2}\nonumber\\
&\times& {\rm Im} \biggl[ (1+ik_1\tau_I)e^{-ik_1\tau_I}H_{n+1/2}^{(1)}(-k_2\tau_I)H_{n+1/2}^{(1)}(-k_3\tau_I)\Biggr.\nonumber\\
&\times&\biggl. \int^{\tau_I} d\tau \tau(1-ik_1\tau)e^{ik_1\tau}H_{n-1/2}^{(2)}(-k_2\tau)H_{n-1/2}^{(2)}(-k_3\tau)\biggr]~,
\eea
\bea
\tilde{\mathcal{I}}_n^{(2)}&=& \frac{\pi^3}{2}\frac{2^{-2n-1}}{\Gamma^2(n+1/2)}(-k_2\tau_I)^{n+1/2}(-k_3\tau_I)^{n+1/2}\nonumber\\
&\times& {\rm Im} \biggl[ (1+ik_1\tau_I)e^{-ik_1\tau_I}H_{n+1/2}^{(1)}(-k_2\tau_I)H_{n+1/2}^{(1)}(-k_3\tau_I)\biggr.\nonumber\\
&\times&\biggl. \int^{\tau_I} d\tau  \tau(1-ik_1\tau)e^{ik_1\tau}H_{n+1/2}^{(2)}(-k_2\tau)H_{n+1/2}^{(2)}(-k_3\tau)\biggr]~.
\eea
Similarly for $ n<-\frac{1}{2}$, 
\bea
\tilde{\mathcal{I}}_n^{(1)}&=& \pi \; \frac{2^{2n} \; \Gamma^2(n+3/2)}{(-k_2\tau_I)^{n+1/2}(-k_3\tau_I)^{n+1/2}}\nonumber\\
&\times& {\rm Im} \biggl[ (1+ik_1\tau_I)e^{-ik_1\tau_I}H_{n+1/2}^{(1)}(-k_2\tau_I)H_{n+1/2}^{(1)}(-k_3\tau_I)\Biggr.\nonumber\\
&\times&\biggl. \int^{\tau_I} d\tau \tau(1-ik_1\tau)e^{ik_1\tau}H_{n-1/2}^{(2)}(-k_2\tau)H_{n-1/2}^{(2)}(-k_3\tau)\biggr]~,\label{J1a}
\eea
\bea
\tilde{\mathcal{I}}_n^{(2)}&=& \pi \; \frac{2^{2n} \; \Gamma^2(n+3/2)}{(-k_2\tau_I)^{n+1/2}(-k_3\tau_I)^{n+1/2}}\nonumber\\
&\times& {\rm Im} \biggl[ (1+ik_1\tau_I)e^{-ik_1\tau_I}H_{n+1/2}^{(1)}(-k_2\tau_I)H_{n+1/2}^{(1)}(-k_3\tau_I)\biggr.\nonumber\\
&\times&\biggl. \int^{\tau_I} d\tau  \tau(1-ik_1\tau)e^{ik_1\tau}H_{n+1/2}^{(2)}(-k_2\tau)H_{n+1/2}^{(2)}(-k_3\tau)\biggr]~.\label{J2a}
\eea

In the limit corresponding to the end of inflation $|k \tau_I| \to 0$, one could show that $\tilde{\mathcal{J}}_n^{(1)} \sim \tilde{\mathcal{I}}_n^{(1)}$ and $\tilde{\mathcal{J}}_n^{(2)} \sim \tilde{\mathcal{I}}_n^{(2)}$. To achieve this result, let's rewrite the integrals (\ref{I1E}) and (\ref{I2E}) as follows,  
\bea
\mathcal{J}_1&=& 2\,{\rm Im} \Biggl[\gamma_{k_1}(\tau_I)A_{k_2}(\tau_I)A_{k_3}(\tau_I)\left(\frac{A'_{k_2}(\tau_I)}{A_{k_2}(\tau_I)}\right)\left(\frac{A'_{k_3}(\tau_I)}{A_{k_3}(\tau_I)}\right)\nonumber \\ 
&\times&\int d\tau  \lambda(\tau) \gamma_{k_1}^*(\tau)A_{k_2}^{'*}(\tau)A_{k_3}^{'*}(\tau)\Biggr],\\
\mathcal{J}_2 &=& 2\,{\rm Im} \Biggl[\gamma_{k_1}(\tau_I)A_{k_2}(\tau_I)A_{k_3}(\tau_I)\left(\frac{A'_{k_2}(\tau_I)}{A_{k_2}(\tau_I)}\right)\left(\frac{A'_{k_3}(\tau_I)}{A_{k_3}(\tau_I)}\right)\nonumber \\ &\times& \int d\tau  \lambda(\tau) \gamma_{k_1}^*(\tau)A_{k_2}^*(\tau)A_{k_3}^*(\tau)\Biggr].
\eea
Using equation (\ref{modfn_A}), the time derivative of mode function can be obtained, 
\bea
A'_k(\tau) = -k A_k(\tau)\; \left[\frac{H^{(1)}_{n-\frac{1}{2}}(-k\tau)}{H^{(1)}_{n+\frac{1}{2}}(-k\tau)}\right]~.\label{modfn_A_derivative}
\eea
By definition, i.e. from (\ref{Int1}), (\ref{Int2}), (\ref{IntJ1}) and (\ref{IntJ2}), we can see that, 
\bea
\frac{{\mathcal{J}}_1}{ {\mathcal{I}}_1
}&=&\frac{|A'^{(0)}_{k_2}(\tau_I)|}{|A^{(0)}_{k_2}(\tau_I)|}\frac{|A'^{(0)}_{k_3}(\tau_I)|}{|A^{(0)}_{k_3}(\tau_I)|} \; \left(\frac{\tilde{\mathcal{J}}_n^{(1)}}{\tilde{\mathcal{I}}_n^{(1)}}\right) \label{J1/I1}\\
\frac{{\mathcal{J}}_2}{ {\mathcal{I}}_2
}&=&\frac{|A'^{(0)}_{k_2}(\tau_I)|}{|A^{(0)}_{k_2}(\tau_I)|}\frac{|A'^{(0)}_{k_3}(\tau_I)|}{|A^{(0)}_{k_3}(\tau_I)|} \; \left(\frac{\tilde{\mathcal{J}}_n^{(2)}}{\tilde{\mathcal{I}}_n^{(2)}}\right).  \label{J2/I2}
\eea
Now, let us take the limit $|k \tau_I| \to 0$ and consider the equations (\ref{J1a}), (\ref{modfn_A_derivative}) and (\ref{J1/I1}),  we will obtain,
\beq
\tilde{\mathcal{J}}_n^{(1)} \sim \tilde{\mathcal{I}}_n^{(1)}. 
\eeq
Similarly by considering the equations (\ref{J2a}), (\ref{modfn_A_derivative}) and (\ref{J2/I2}),
\beq
\tilde{\mathcal{J}}_n^{(2)} \sim \tilde{\mathcal{I}}_n^{(2)}. 
\eeq

One can evaluate these integrals for different values of $n$ which are listed in the appendix of \cite{Jain:2012vm}. For completeness, we present the result below for the most interesting case of $n=2$ which has been used to plot the non-linearity parameters in Sec. \ref{nl-param}, as
\bea
\tilde{\mathcal{I}}_2^{(1)}&=& \frac{-1}{(k_2 k_3)^{3/2} k_t^2} \nonumber \\
&\times&\left[-k_1^3 - 2 k_1^2 (k_2 + k_3) - 
 2 k_1 (k_2^2 + k_2 k_3 + k_3^2) - (k_2 + k_3) (k_2^2 + k_2 k_3 + 
    k_3^2)\right]
\eea
and 
\bea
\tilde{\mathcal{I}}_2^{(2)}&=& \frac{-1}{(k_2 k_3)^
  {5/2} k_t^2} \nonumber \\
   &\times&\left[(k_1 + k_2)^2 (-3 k_1^3 - 3 k_1^2 k_2 - 
      k_2^3) + (k_1 + k_2) (-9 k_1^3 - 6 k_1^2 k_2 - 
      2 k_2^3) k_3\right.  \nonumber \\
      &+& (-9 k_1^3 - 6 k_1^2 k_2 - 
      2 k_1 k_2^2 - 2 k_2^3) k_3^2 \nonumber \\
      & -&\left. 2 (2 k_1^2 + k_1 k_2 + k_2^2) k_3^3 - 
   2 (k_1 + k_2) k_3^4 - k_3^5 + 
   3 k_1^3 k_t^2 (\gamma+\ln(-k_t\tau_I) )\right]
\eea
where $k_t =k_1+k_2+k_3$ and $\gamma$ is the Euler gamma constant. 


\section{Validity of the semi-classical approach}
\label{validity-sca}

Here, we shall quickly review the validity of the semi-classical approach for the three-point correlators involving tensor perturbation. For the semi-classical derivation of single field consistency relations, one argues that the effect of the long wavelength metric perturbation can be absorbed into a local coordinate transformation. Then the two point functions in a modified background can be calculated by rescaling the coordinates. And use them to calculate the squeezed limit three-point functions. We can see that the same arguments will be valid also in our scenario. So let's try to establish this fact at the action level. To start with, the gauge field action of our interest is given by 
\bea
S_{EM}=-\frac{1}{4} \int d^4 x \sqrt{-g} \,\lambda(\phi)F_{\mu \nu}F^{\mu \nu}.
\eea
Now we can calculate the leading order action by introducing tensor perturbation, 
\bea
\label{Pert_Action}
S=S^{(0)}-\frac{1}{4}\int d^4 x \lambda(\phi) \left[2F_{0i}F_{0j}\gamma^{ij}-2F_{ij}F_{im}\gamma^{mj}\right],
\eea
where $\gamma_{ij}$ is the tensor perturbation and $S^{(0)}$ denotes the  free part of the action. So the second term in RHS of equation (\ref{Pert_Action}) corresponds to interacting part of the action. One can express this interacting part in the Fourier space and study its squeezed limit. In the squeezed limit, wherein the momentum corresponding to $\gamma_{ij}$ is much smaller than the gauge fields, we can write, 
\bea
S=S^{(0)}-\frac{1}{2}\int d\tau \lambda(\tau) \int \frac{d^3{\bf k}}{(2\pi)^3} \Big[A'_i({\bf k}) A'_j(-{\bf k})\gamma_{ij}-k^2 A_i({\bf k}) A_j(-{\bf k})\gamma_{ij}-k_i k_j \gamma_{ij}A^2\Big]\label{perturbedh}~.
\eea
Since $\gamma_{ij}$ is time independent in the squeezed limit, we can use the on-shell conditions to simplify equation (\ref{perturbedh}) as follows, 
\bea
\int d\tau A'_i({\bf k}) A'_j(-{\bf k}) &=& \int d\tau \left( \frac{\partial}{\partial \tau}(A_i \lambda(\tau)A'_j)-A_i \lambda(\tau)A''_j-A_i \lambda '(\tau)A'_j\right) 
\nonumber \\
&=&-\int d\tau \lambda(\tau)A_i  \left( A''_j+\frac{ \lambda'}{\lambda}A'_j\right)\nonumber \\
&=&\int d\tau \lambda(\tau)k^2 A_i({\bf k}) A_j(-{\bf k})\label{onshel}~,
\eea
where we have used the equation of motion of $A_i$ in the last line of the above equation. Substitute equation (\ref{onshel}) into (\ref{perturbedh}) and simplify, 
\bea
S=S^{(0)}+\frac{1}{2}\int d\tau \lambda(\tau) \int \frac{d^3{\bf k}}{(2\pi)^3} k_i k_j \gamma_{ij}A^2~.
\eea
$S^{(0)}$ can be rewritten as, 
\bea
S^{(0)}=\frac{1}{2}\int d\tau \lambda(\tau) \int \frac{d^3{\bf k}}{(2\pi)^3}\left((A'_i)^2-k^2 A^2\right)~.
\eea
Then the total action becomes, 
\bea
S &=& \frac{1}{2} \int d\tau \lambda(\tau) \int \frac{d^3{\bf k}}{(2\pi)^3}\left((A'_i)^2-(k^2-k_i k_j \gamma_{ij}) A^2\right) \nonumber \\
&=& \frac{1}{2} \int d\tau \lambda(\tau) \int \frac{d^3{\bf k}}{(2\pi)^3}\left((A'_i)^2-\tilde{k}^2 A^2\right)~,
\eea
where $\tilde{k}^2=k^2-k_i k_j \gamma_{ij}$. One can identify this as the rescaled momentum in the rescaled background, i.e.,
$
\label{rescaledBG}
ds^2 =-dt^2+a^2(t) d \tilde{x}^2
$
with $ d\tilde x^2 \to  dx^2+\gamma^B_{ij}dx^idx^j$. So this establishes the fact that the effect of the long wavelength graviton mode can be absorbed into a local coordinate transformation. 


\section{The in-in formalism and the dynamical correction term}
\label{dynamical}
The electric field is defined through the time-derivative of the vector potential. When promoting these to operators and taking expectation values, we need to be careful and distinguish between time-derivatives of operators in the Heisenberg picture and the interaction picture. The relation between the electric field and the time-derivative of the vector potential is simplest in the Heisenberg picture where only the operators depend on time. We will therefore start by considering the time derivative of an operator in the Heisenberg picture, and then translate that into the corresponding operator in the interaction picture where also the state depends on time. In this appendix we will carefully distinguish between interaction picture and Heisenberg picture operators, but in the main text all discussion is in the interaction picture and all operators are implicitly understood to be in the interaction picture unless explicitly defined otherwise.

Let $\mathcal{O} (\tau)$ be an observable of our interest which is constructed out of field operators and their derivatives. The time evolution of this observable is given by the Heisenberg equation of motion 
\bea 
\label{HEM}
\frac{d}{d\tau}\mathcal{O}^H (\tau)=i[H_{\rm tot}(\tau),\mathcal{O}^H (\tau)].
\eea
One can use the standard procedure in quantum field theory and express the solution in terms of a unitary operator $U$ as 
\bea
\mathcal{O}^H (\tau)=U^{-1}(\tau, \tau_i)\mathcal{O}^H (\tau_i)U(\tau, \tau_i),
\eea
where $\tau_i$ denotes the initial time and $U$ obeys the following equation 
\bea 
\label{eq_unitary_op}
i \frac{\p}{\p\tau}U(\tau, \tau_i)=H_{\rm tot}(\tau)U(\tau, \tau_i),
\eea
and satisfies the following properties
\bea 
U(\tau, \tau)=1, \qquad U^{-1}(\tau, \tau_i)U(\tau, \tau_i)=1, \qquad U(\tau_3,\tau_2)U(\tau_2, \tau_1)=U(\tau_3, \tau_1).
\eea
Thus, $U$ can be interpreted as a time evolution operator which evolves the operator $\mathcal{O}$ from an initial time  $\tau_i$ to a later time $\tau$. 

In this work, we are interested in the expectation value of $\mathcal{O} (\tau)$, i.e.,
\bea
\label{exp_value}
\lle \mathcal{O} (\tau)\rgr \equiv \lle in \rb \mathcal{O}^H (\tau)\lb in\rgr=\lle in \rb U^{-1}(\tau, \tau_i)\mathcal{O}^H (\tau_i)U(\tau, \tau_i) \lb in\rgr,
\eea
where $ | in \rangle$ denotes the initial quantum state of the system. Since we are working in the Heisenberg picture, the states are independent of time. 
Now, our main goal is to calculate equation (\ref{exp_value}). In order to do so, one has to split the total Hamiltonian as follows
\bea 
H_{\rm tot}(\tau)=H_0(\tau)+H_{\text{int}}(\tau),
\eea
where $H_0(\tau)$  and $H_{\text{int}}(\tau)$ denote the free and the interacting parts of the Hamiltonian, respectively. Since we know the evolution of the system under free Hamiltonian, i.e. without any interaction, we can make use of it to understand the time evolution under full Hamiltonian. This can be easily illustrated by introducing the interaction picture. 
To this end, let us define $U_0(\tau, \tau_i)$ as 
\bea
\label{free_U}
i \frac{\p}{\p\tau}U_0(\tau, \tau_i)=H_{0}(\tau)U_0(\tau, \tau_i),
\eea
so one can rewrite the expectation value of $\mathcal{O} (\tau)$ as in equation (\ref{exp_value}) as follows
\bea 
\lle in \rb U^{-1}(\tau, \tau_i)\mathcal{O}^H (\tau_i)U(\tau, \tau_i)\lb in\rgr=\lle in \rb F^{-1}(\tau, \tau_i)U_0^{-1}(\tau, \tau_i)\mathcal{O}^H (\tau_i)U_0(\tau, \tau_i)F(\tau, \tau_i)\lb in\rgr,
\eea
with 
\bea 
F(\tau, \tau_i)=U_0^{-1}(\tau, \tau_i)U(\tau, \tau_i).
\eea
Then, we can define the operator $ \mathcal{O} $ in the interaction picture in the following manner as 
\bea 
\label{interaction_pic}
\mathcal{O}^{I} (\tau)=U_0^{-1}(\tau, \tau_i)\mathcal{O}^H (\tau_i)U_0(\tau, \tau_i).
\eea
Using this definition, we can now express the expectation value of $\mathcal{O} (\tau)$ as
\bea 
\lle \mathcal{O} (\tau)\rgr=\lle in \rb F^{-1}(\tau, \tau_i)\mathcal{O}^{I} (\tau)F(\tau, \tau_i)\lb in\rgr,
\eea
with 
\bea 
\label{eq.F}
i \frac{\p}{\p\tau}F(\tau, \tau_i)=H^I_{\text{int}}(\tau)F(\tau, \tau_i),
\eea
and the superscript $I$ refers to the interaction picture. The solution to the above equation can be written in terms of $H^I_{\text{int}}$ as
\bea 
F(\tau, \tau_i)=T \left(\text{exp}\left[-i\int_{\tau_i}^{\tau}d\tau'H^I_{\text{int}}(\tau')\right]\right),
\eea
where $T$ is the time ordering operator. Using this equation, one can calculate the expectation value of an operator $\mathcal{O}$  at time $\tau$, 
\bea 
\lle \mathcal{O} (\tau)\rgr=\lle in \rb \bar{T}\left(e^{i\int_{\tau_i}^{\tau}d\tau'H^I_{\text{int}}(\tau')}\right)\mathcal{O}^{I} (\tau)T\left(e^{-i\int_{\tau_i}^{\tau}d\tau'H^I_{\text{int}}(\tau')}\right)\lb in\rgr.
\eea
This result is the master formula of the in-in formalism for calculating equal time correlation functions. It was first developed by Schwinger \cite{Schwinger:1960qe} and others \cite{Mahanthappa:1962ex,Bakshi:1963bn,Keldysh:1964ud,Chou:1984es} and then later applied to cosmology \cite{Jordan:1986ug,PhysRevD.35.495}.

In this paper, we are interested in the correlators $\gamma A_{\mu}A^{\mu}$, $\gamma B_{\mu}B^{\mu}$ and $\gamma E_{\mu}E^{\mu}$. The construction of interaction picture of $\gamma A_{\mu}A^{\mu}$ and $\gamma B_{\mu}B^{\mu}$ is trivial but $\gamma E_{\mu}E^{\mu}$ requires more attention, as it involves the time derivative of the vector potential. Let us therefore compare the time-derivative of the operator in the Heisenberg picture and in the interaction picture.

The time derivative of the field has a clear meaning as only the operators evolve in time in the Heisenberg picture,
\beq
\frac{d}{d\tau}\lle \mathcal{O} (\tau)\rgr= \lle in \rb \frac{d}{d\tau} \mathcal{O}^H (\tau)\lb in\rgr~.
\eeq
Now, identifying $\mathcal{Q}^H(\tau) \equiv (d/d\tau)\mathcal{O}^H (\tau)$ and using (\ref{interaction_pic}), we have $\mathcal{Q}^I(\tau)=U_0^{-1}(\tau,\tau_i)\mathcal{Q}^H (\tau_i)U_0(\tau,\tau_i)$. Using equation (\ref{HEM}), we get
\bea 
\mathcal{Q}^I(\tau)= U_0^{-1}(\tau, \tau_i)\left(\frac{d}{d\tau} \mathcal{O}^H(\tau)\right)\bigg|_{\tau=\tau_i}U_0(\tau, \tau_i)&=& iU_0^{-1}(\tau, \tau_i)[H_{\rm tot}(\tau_i),\mathcal{O}^H(\tau_i)]U_0(\tau, \tau_i) \nonumber \\
&=&iU_0^{-1}(\tau, \tau_i)[H_{0}(\tau_i)+H_{\rm int}(\tau_i),\mathcal{O}^H(\tau_i)]U_0(\tau, \tau_i)\nonumber \\ \label{A_int}
&=&i[H_{0}(\tau),\mathcal{O}^I(\tau)]+i[H_{\rm int}^I(\tau),\mathcal{O}^I(\tau)] \nonumber \\
&=&\frac{d}{d\tau}\mathcal{O}^I(\tau)+i[H_{\rm int}^I(\tau),\mathcal{O}^I(\tau)]  .
\eea 
We conclude, that if we have an operator, which in the Heisenberg picture would take the form of a time derivative of an operator, i.e.  $\mathcal{Q}^H(\tau) \equiv (d/d\tau)\mathcal{O}^H (\tau)$, then the interaction picture version of that operator is the time derivative of the interaction picture operator plus and extra piece, i.e. $\mathcal{Q}^I(\tau)=(d/d\tau)\mathcal{O}^I(\tau)+i[H_{\rm int}^I(\tau),\mathcal{O}^I(\tau)]$.

Applying this observation to the electric field in the interaction picture $E_{\mu}^I$, we find
\bea 
\label{app_E}
E^I_i(\tau)=\frac{1}{a(\tau)}\left[\left(\frac{d}{d\tau}A_i^I(\tau)\right)+i[H_{\rm int}^I(\tau),A_i^I(\tau)]\right].
\eea
Since all the operators in the above equation are in the interaction picture, let's suppress the corresponding superscript $I$ for brevity. Then one can express the observable $\gamma E_{\mu}E^{\mu}$, to the leading order, in the interaction picture as follows, 
\bea
 \gamma E_{\mu}E^{\mu}=\frac{1}{a^4} \left[\gamma \frac{dA_i}{d\tau}\frac{dA_i}{d\tau}+i \gamma \left(\frac{dA_i}{d\tau}[H_{\rm int},A_i]+[H_{\rm int},A_i]\frac{dA_i}{d\tau}\right)-\gamma \gamma_{ij} \frac{dA_i}{d\tau}\; \frac{dA_j}{d\tau}\right].
\eea
We refer to the second term as the dynamical correction term and the third as the kinematical correction term. 

\bibliography{inf-mf-ng.bib}
\bibliographystyle{JHEP}

\end{document}